\documentclass[twoside,twocolumn,9pt]{article}
\usepackage{extsizes}
\usepackage[super,sort&compress,comma]{natbib} 
\usepackage[version=3]{mhchem}
\usepackage[left=1.5cm, right=1.5cm, top=1.785cm, bottom=2.0cm]{geometry}
\usepackage{balance}
\usepackage{mathptmx}
\usepackage{sectsty}
\usepackage{graphicx} 
\usepackage{lastpage}
\usepackage[format=plain,justification=justified,singlelinecheck=false,font={stretch=1.125,small,sf},labelfont=bf,labelsep=space]{caption}
\usepackage{float}
\usepackage{fancyhdr}
\usepackage{fnpos}
\usepackage[english]{babel}
\addto{\captionsenglish}{%
  
}
\usepackage{array}
\usepackage{droidsans}
\usepackage{charter}
\usepackage[T1]{fontenc}
\usepackage[usenames,dvipsnames]{xcolor}
\usepackage{setspace}
\usepackage[compact]{titlesec}

\usepackage{amsmath, amssymb}

\definecolor{darkblue}{rgb}{0,0,0.6}
\definecolor{darkred}{rgb}{0.6,0,0}
\usepackage[colorlinks=true,urlcolor=darkblue,citecolor=darkblue, linkcolor=darkred,hyperfootnotes=false]{hyperref}

\usepackage{amsmath}
\usepackage{siunitx}
\usepackage{mathtools}
\usepackage{subcaption}

\usepackage{epstopdf}

\definecolor{cream}{RGB}{222,217,201}

\begin{document}

\pagestyle{fancy}
\thispagestyle{plain}
\fancypagestyle{plain}{
\renewcommand{\headrulewidth}{0pt}
}

\makeFNbottom
\makeatletter
\renewcommand\LARGE{\@setfontsize\LARGE{15pt}{17}}
\renewcommand\Large{\@setfontsize\Large{12pt}{14}}
\renewcommand\large{\@setfontsize\large{10pt}{12}}
\renewcommand\footnotesize{\@setfontsize\footnotesize{7pt}{10}}
\makeatother

\renewcommand{\thefootnote}{\fnsymbol{footnote}}
\renewcommand\footnoterule{\vspace*{1pt}%
\color{cream}\hrule width 3.5in height 0.4pt \color{black}\vspace*{5pt}} 
\setcounter{secnumdepth}{5}

\makeatletter 
\renewcommand\@biblabel[1]{#1}            
\renewcommand\@makefntext[1]%
{\noindent\makebox[0pt][r]{\@thefnmark\,}#1}
\makeatother 
\renewcommand{\figurename}{\small{Fig.}~}
\sectionfont{\sffamily\Large}
\subsectionfont{\normalsize}
\subsubsectionfont{\bf}
\setstretch{1.125} 
\setlength{\skip\footins}{0.8cm}
\setlength{\footnotesep}{0.25cm}
\setlength{\jot}{10pt}
\titlespacing*{\section}{0pt}{4pt}{4pt}
\titlespacing*{\subsection}{0pt}{15pt}{1pt}

\fancyfoot{}
\fancyfoot[LO,RE]{\vspace{-7.1pt}\includegraphics[height=9pt]{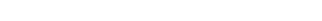}}
\fancyfoot[CO]{\vspace{-7.1pt}\hspace{13.2cm}\includegraphics{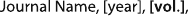}}
\fancyfoot[CE]{\vspace{-7.2pt}\hspace{-14.2cm}\includegraphics{head_foot/RF}}
\fancyfoot[RO]{\footnotesize{\sffamily{1--\pageref{LastPage} ~\textbar  \hspace{2pt}\thepage}}}
\fancyfoot[LE]{\footnotesize{\sffamily{\thepage~\textbar\hspace{3.45cm} 1--\pageref{LastPage}}}}
\fancyhead{}
\renewcommand{\headrulewidth}{0pt} 
\renewcommand{\footrulewidth}{0pt}
\setlength{\arrayrulewidth}{1pt}
\setlength{\columnsep}{6.5mm}
\setlength\bibsep{1pt}

\makeatletter 
\newlength{\figrulesep} 
\setlength{\figrulesep}{0.5\textfloatsep} 

\newcommand{\topfigrule}{\vspace*{-1pt}%
\noindent{\color{cream}\rule[-\figrulesep]{\columnwidth}{1.5pt}} }

\newcommand{\botfigrule}{\vspace*{-2pt}%
\noindent{\color{cream}\rule[\figrulesep]{\columnwidth}{1.5pt}} }

\newcommand{\dblfigrule}{\vspace*{-1pt}%
\noindent{\color{cream}\rule[-\figrulesep]{\textwidth}{1.5pt}} }

\makeatother

\renewcommand{\boldsymbol}[1]{{\bf #1}}
\DeclareSIUnit{\Vpp}{V_{pp}}

\newcommand{\cA}{\mathcal{A}}
\newcommand{\cB}{\mathcal{B}}
\newcommand{\cC}{\mathcal{C}}
\newcommand{\cD}{\mathcal{D}}
\newcommand{\cE}{\mathcal{E}}
\newcommand{\cF}{\mathcal{F}}
\newcommand{\dF}{\mathscr{F}}
\newcommand{\cG}{\mathcal{G}}
\newcommand{\cH}{\mathcal{H}}
\newcommand{\cJ}{\mathcal{J}}
\newcommand{\cK}{\mathcal{K}}
\newcommand{\mM}{\mathbb{M}}
\newcommand{\cO}{\mathcal{O}}
\newcommand{\cP}{\mathcal{P}}
\newcommand{\dP}{\mathcal{P}}
\newcommand{\cQ}{\mathcal{Q}}
\newcommand{\cR}{\mathcal{R}}
\newcommand{\cU}{\mathcal{U}}
\newcommand{\cW}{\mathcal{W}}
\newcommand{\cZ}{\mathcal{Z}}
\newcommand{\Cinf}{C_{\infty}}
\newcommand{\dd}{\text{d}}
\newcommand{\ii}{\text{i}}
\newcommand{\p}{\partial}

\newcommand{\fig}{{Figure~}}
\newcommand{\iden}{{\bf 1}}
\newcommand{\ind}[1]{_{\mathrm{#1}}}
\renewcommand{\AA}{\boldsymbol{A}}
\newcommand{\mcA}{\mathcal{A}}
\newcommand{\mcC}{\mathcal{C}}
\newcommand{\ed}{\mathrm{e}}
\newcommand{\ee}{\boldsymbol{e}}
\newcommand{\EE}{\boldsymbol{E}}
\newcommand{\ff}{\boldsymbol{f}}
\newcommand{\FF}{\boldsymbol{F}}
\newcommand{\GG}{\boldsymbol{G}}
\newcommand{\id}{\mathrm{i}}
\newcommand{\jj}{\boldsymbol{j}}
\newcommand{\JJ}{\boldsymbol{J}}
\newcommand{\kk}{\boldsymbol{k}}
\newcommand{\nn}{\boldsymbol{n}}
\newcommand{\mcO}{\mathcal{O}}
\newcommand{\PP}{\boldsymbol{P}}
\newcommand{\qq}{\boldsymbol{q}}
\newcommand{\rr}{\boldsymbol{r}}
\renewcommand{\ss}{\boldsymbol{s}}
\newcommand{\TT}{\boldsymbol{T}}
\newcommand{\mcV}{\mathcal{V}}
\newcommand{\uu}{\boldsymbol{u}}
\newcommand{\vv}{\boldsymbol{v}}
\newcommand{\xx}{\boldsymbol{x}}
\newcommand{\XX}{\boldsymbol{X}}
\newcommand{\yy}{\boldsymbol{y}}
\newcommand{\zz}{\boldsymbol{z}}
\newcommand{\eex}{\hat{\boldsymbol{e}}_x}
\newcommand{\eeta}{\boldsymbol{\eta}}
\newcommand{\eepsilon}{\boldsymbol{\varepsilon}}
\newcommand{\pphi}{\boldsymbol{\phi}}
\newcommand{\PPhi}{\boldsymbol{\Phi}}
\newcommand{\ppsi}{\boldsymbol{\psi}}
\newcommand{\xxi}{\boldsymbol{\xi}}
\newcommand{\XXi}{\boldsymbol{\Xi}}
\newcommand{\ssigma}{\boldsymbol{\sigma}}
\newcommand{\zzeta}{\boldsymbol{\zeta}}
\newcommand{\zzero}{\boldsymbol{0}}
\newcommand{\Pe}{\mathrm{Pe}}
\newcommand{\red}[1]{\textcolor{BrickRed}{#1}}
\newcommand{\blue}[1]{\textcolor{NavyBlue}{#1}}

\twocolumn[
  \begin{@twocolumnfalse}
{\includegraphics[height=30pt]{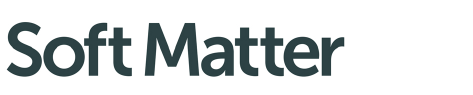}\hfill\raisebox{0pt}[0pt][0pt]{\includegraphics[height=55pt]{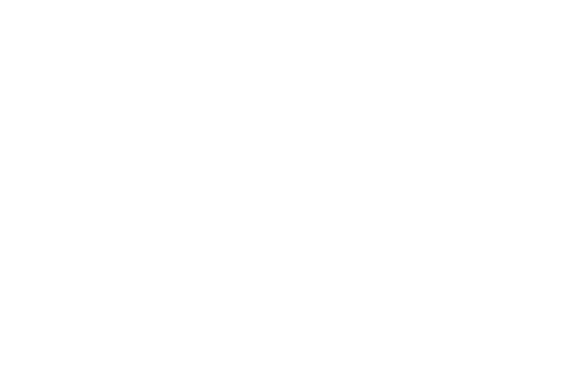}}\\[1ex]
\includegraphics[width=18.5cm]{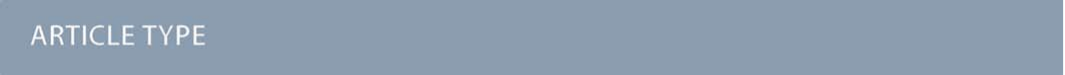}}\par
\vspace{1em}
\sffamily
\begin{tabular}{m{4.5cm} p{13.5cm} }

\includegraphics{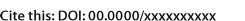} & \noindent\LARGE{\textbf{Learning general pair interactions between \newline self-propelled particles \textit{$^{\dag}$} }} \\
\vspace{0.3cm} & \vspace{0.3cm} \\

 & \noindent\large{Jérôme Hem,\textit{$^{ab}$} Alexis Poncet,\textit{$^{c}$} Pierre Ronceray, \textit{$^{d}$}  Daiki Nishiguchi, \textit{$^{ef}$} and Vincent Démery$^{\ast}$\textit{$^{ac}$}} \\

\includegraphics{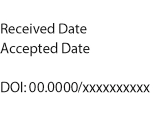} & \noindent\normalsize{Synthetic active matter systems, such as active colloids, often have complex interactions, which can be of hydrodynamic, chemical or electrostatic origin and cannot be computed from first principles.
Here, we use Stochastic Force Inference to learn general pair interactions, including transverse forces and torques, between self-propelled Janus particles from experimental trajectories. 
We use data from two experiments: one where the particles flock, and one where the system remains disordered.
The learned interactions are then fed to numerical simulations, which reproduce all the experimental observables and could be extrapolated to different densities.
Overall, we find that the radial interaction is mostly repulsive and isotropic, while the angular interaction has a richer angular dependence, which controls the behavior of the system; the transverse interaction is negligible.
Finally, testing the symmetry relations obeyed by the inferred interactions allows us to show that they cannot come from electrostatics only, so that they must have a hydrodynamic component.} \\

\end{tabular}

 \end{@twocolumnfalse} \vspace{0.6cm}

  ]

\renewcommand*\rmdefault{bch}\normalfont\upshape
\rmfamily
\section*{}
\vspace{-1cm}


\footnotetext{\textit{$^{a}$~Gulliver UMR CNRS 7083, ESPCI Paris, Université PSL, 75005 Paris, France}}
\footnotetext{\textit{$^{b}$~Institut Jean Lamour, CNRS, Université de Lorraine, 54011 Nancy, France}}
\footnotetext{\textit{$^{c}$~Univ Lyon, ENS de Lyon, Univ Claude Bernard Lyon 1, CNRS, Laboratoire de Physique, \newline F-69342 Lyon, France }}
\footnotetext{\textit{$^{d}$~Aix Marseille Univ, CNRS, CINaM (UMR 7325), Turing Centre for Living Systems \newline (CENTURI), Marseille, France }}
\footnotetext{\textit{$^{e}$~Department of Phsyics, Institute of Science Tokyo, Ookayama 2-12-1, \newline Tokyo, 152-8551, Japan}}
\footnotetext{\textit{$^{f}$~Department of Physics, The University of Tokyo, Hongo 7-3-1, \newline Tokyo, 113-0033, Japan }}
\footnotetext{\textit{$^{\ast}$~E-mail: vincent.demery@espci.psl.eu}}

\footnotetext{\dag~The program codes are available in the public repository: \newline \url{https://github.com/jhem272/sfiabp_py}.}



\section{Introduction}
\label{Sec:Intro}


The field of active matter has emerged as a framework to describe living systems as diverse as bacterial colonies~\cite{Aranson2022, Xu2023Autonomous}, schools of animals~\cite{Vicsek1995, Gautrais2012, Chen2023} or human crowds~\cite{Bain2019, Cordes2024Dimensionless, Gu2025}.
To help the development of new models, it was soon realized that synthetic active matter systems were needed to perform better controlled experiments~\cite{Bechinger2016}.
Focusing on self-propelled polar particles, experimental realizations have already used vibrated disks~\cite{Deseigne2010,Weber2013}, swimming droplets~\cite{Izri2014, Hokmabad2022Chemotactic, Lin2024Emergent}, as well as rolling and suspended colloids~\cite{Bricard2013, Palacci2010}. 
In particular, Janus particles propelled by a vertical electric field through induced-charge electrophoresis~\cite{Bazant2010, Boymelgreen2016, Nishiguchi2015} are very attractive model systems, with rich collective behaviors that depend on tunable pair interactions~\cite{Yan2016Reconfiguring, Nishiguchi2018, Zhang2021, Poncet2021, Iwasawa2021, Nishiguchi2023Deciphering, Das2024Flocking}. However, despite being well controlled, computing the interactions between the particles in these systems is very challenging.
For example, in the case of Janus particles propelled by an electric field, hydrodynamic interactions were sometimes neglected and only electrostatic interactions were taken into account~\cite{Yan2016Reconfiguring,Zhang2021,Das2024Flocking}, but these can only be computed under several approximations, among which is the hypothesis that the two colloids are far from any boundary.
Considering that the particles are actually in the vicinity of an electrode, this is a rather crude assumption.
Similarly, the hydrodynamic flow around a particle can be computed from the Stokes equation, assuming that the particle is far from any other particle and far from the electrodes~\cite{Nishiguchi2015}.
If the interactions cannot be computed but their knowledge is essential to understand the behavior of the system, they need to be measured in experiments. 
This approach is routinely used for the self-propulsion velocity $U$, which is always measured in experiments in the dilute regime but never computed from first principles.
Although the methods to extract one particle parameters such as $U$ are usually quite straightforward, a mean value obtained from a distribution, they cannot be readily applied to measure the interactions because of the many supplementary degrees of freedom.
Instead, one needs to develop more sophisticated methods that have the ability to measure the interactions in the presence of multiple source of noises, which are always present in experiments on colloidal systems.    


To measure the interactions from experimental trajectories, many approaches can be followed. 
One is Machine Learning (ML) based techniques, where the interactions are encoded in a neural network~\cite{Cichos2020}, which gained in popularity over the last years.
Already several attempts have been made for colloidal systems:
the radial pair interaction between colloids has been measured in Ref.~\cite{RuizGarcia2024Discovering}, whereas  the coarse-grained hydrodynamical model has been inferred directly in Refs.~\cite{Joshi2022, Supekar2023Learning}. 
One can also find model-free ML techniques whose goal is to forecast the dynamics.
For example the authors in Ref.~\cite{Ha2021Unraveling} successfully predicted the dynamics of complex particle assemblies up to few seconds. 
Besides, it is worth mentioning the existence of other promising frameworks, based on ML techniques or not, which focus on the inference of specific dynamical quantities, such as memory kernels for the generalized Langevin equation~\cite{Jung2017} or specific potentials~\cite{Malpica2023, Torquato2022}. 

In this article, we choose another approach called Stochastic Force Inference (SFI)~\cite{Frishman2020}.
This versatile and ML-free framework relies on a robust denoising process, based on Ito-Stratonovich conversion, that manage to separate the deterministic force from the intrinsic and the measurement noises. 
It is then followed by a standard fitting procedure where the displacements and the diffusion are decomposed on a basis.
This robustness against different source of noises makes it specifically relevant for the study of experimental colloidal systems. 
To date, in the field of soft matter, this approach has been used only to infer the underdamped dynamic of a single particle~\cite{Bruckner2020} although, in principle, it can be applied to large assemblies. 
For example, Ref.~\cite{Frishman2020} demonstrates the inference of isotropic radial pair interactions  between simulated active polar particles.

\begin{figure*}[t]
\begin{center}
\includegraphics[width=170mm]{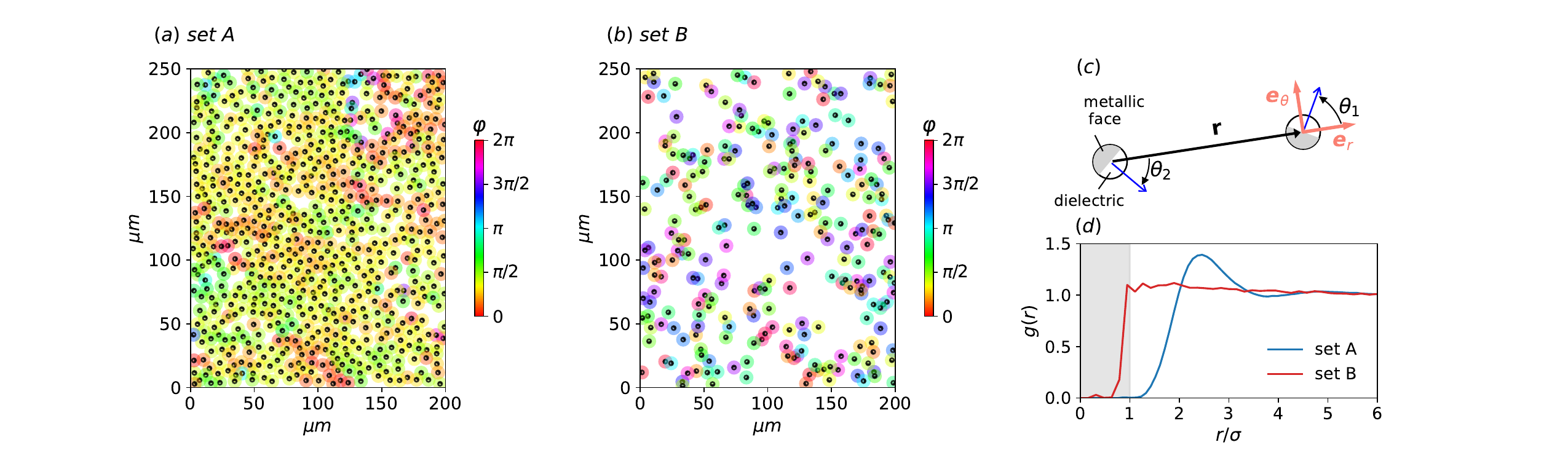}
\end{center}
\caption{ (a,b) Snapshots of size $\qty{200}\times\qty{250}{\micro m^2}$ for the two sets $A$ and $B$. The colored rings surrounding the particles indicate the value of the polar angle $\varphi\in[0,2\pi]$. The direction $\varphi=0$ is along the horizontal axis. Note that for set A, the particles move with the metallic (dark) face ahead whereas for set B, they move with the dielectric (white) face ahead.
    (c) Notations for the relative distance and orientations between particles $1$ and $2$ moving with the white face ahead; the blue arrows indicate the direction of the self-propulsion.
    (d) Radial pair correlation function $g(r)$ for the two sets $A$ and $B$.
    The gray area indicates the range $r/\sigma<1$ with the particle diameter $\sigma=\qty{3.17}{\micro m}$ in both cases.}
\label{fig:schematic}
\end{figure*}

Here, we use SFI to determine the most general pair interactions between Janus particles self-propelled by an AC electric field, from the experimental trajectories, for two different experiments~\cite{Iwasawa2021, Poncet2021}.
The first one, denoted set $A$ in what follows, contains a rather high density of particles (area fraction $\phi_A=\num{9.7e-2}$), which form a flock
with true long range polar order and giant number fluctuations~\cite{Iwasawa2021} (Fig.~\ref{fig:schematic}(a)).
In this set, the high density of particles makes the inference challenging, as it requires to disentangle the forces from the different sources of noise and between many particles.
In contrast, the second dataset, denoted set $B$, contains a lower density of particles ($\phi_B=\num{4e-2}$) with weaker interactions, and the system remains in a disordered state~\cite{Poncet2021} (Fig.~\ref{fig:schematic}(b)).
This is evidenced either in a movie, where many collisions between the particles can be observed, or in the radial pair correlation function $g(r)$, where particles are observed in close contact (Fig.~\ref{fig:schematic}(d)). 
Here, the inference is challenging because of the weak interactions that take place at short distances between the particles. 

This article is structured as follows. 
In part~\ref{Sec:Method}, we define the model and the interactions to consider for the self-propelled particles.
Then, we explain the functionnal basis used to decompose the forces.
In part~\ref{Sec:Flocking-System} and part~\ref{Sec:Disordered-System}, we run SFI and discuss the inferred pair interactions, respectively for the flocking and the disordered experiments. In both cases, the quality of the inference is assessed by running numerical simulations with the inferred interactions, so as to reproduce quantitatively the experimental observables. 
Specific to the flocking system, we show in addition that it is possible to extrapolate the dynamic at different densities, which could be useful for example to predict the transition between the disordered and the ordered states. 
Finally in part~\ref{Sec:Origin}, we discuss the possible origins of the interaction.
The basis used to describe the interaction is chosen to satisfy general symmetry relations.
If the interactions were purely electrostatic, they would derive from a potential and several additional symmetry relations would be satisfied.
In the flocking experiment, we show that the first order harmonic terms of the interactions can be explained by a dominant electrostatic interaction but that the higher order terms cannot, which proves that another contribution, probably of hydrodynamic origin, should be present, supporting the flocking mechanisms discussed in Ref.~\cite{Iwasawa2021}. 
For the disordered experiment, the lack of symmetry relations at any order suggests instead a strong hydrodynamic component compared to the electrostatic one. 

\section{Method}
\label{Sec:Method}

\subsection{Model}

Modeling and determining general interactions between the particles is challenging, mainly because of the many degrees of freedom. 
However, the range of possibilities can be drastically reduced if we assume the particles identical and consider only pairwise interactions.
Considering particles moving in two dimensions, these assumptions reduce the effect of the particle 2 on the particle 1 to the presence of an additional velocity $\vv$ and an angular velocity $\omega$. 
These velocities enter in the overdamped equations of motion for the position $\rr_i(t)$ and orientation $\varphi_i(t)$ of the particle $i$ as:
\begin{align}
  \dot \rr_i (t) &=U\ee_{\varphi_i(t)}+\sum_{j\neq i}\vv(\rr_i-\rr_j,\varphi_i,\varphi_j)+\sqrt{2D}\eeta_i(t),\\
  \dot \varphi_i(t) &=\sum_{j\neq i}\omega(\rr_i-\rr_j,\varphi_i,\varphi_j)+\sqrt{2D_r}\xi_i(t),
\end{align}
where $U$ is the self-propulsion velocity, $D$, $D_r$ are the translational and rotational diffusion coefficients, $\eeta_i(t)$ and $\xi_i(t)$ are Gaussian white noises with unit variance.
Note that the term ``velocity'', which is used for $U$, $\vv$ and $\omega$, corresponds to the ``drift'' in the framework of Stochastic Differential Equations.
The interaction velocity $\vv$ and angular velocity $\omega$ are the products of the translational and angular mobilities with the interaction force and torque; the velocities are the natural outcome of the analysis of the trajectories. 

Further assuming the rotational invariance of the system, these quantities depend on the distance $r=|\rr|$ between the particles, with $\rr=\rr_i-\rr_j$, and on their orientations $\theta_i$ and $\theta_j$ with respect to $\rr$ (Fig.~\ref{fig:schematic}(c)).
Last, the additional velocity $\vv$ can be decomposed in a component along $\rr$, $v_r$, and a component perpendicular to it, $v_\theta$:
\begin{equation}
  \vv(r,\theta_1,\theta_2) = v_r (r,\theta_1,\theta_2) \ee_r + v_\theta (r,\theta_1,\theta_2)\ee_\theta.
\end{equation}
Altogether, the interactions are encoded in three scalar functions of three scalar variables, $v_r (r,\theta_1,\theta_2)$, $v_\theta (r,\theta_1,\theta_2)$ and $\omega (r,\theta_1,\theta_2)$.

\subsection{Fit functions}

We now define the basis on which to decompose the particle velocities. 
For one particle forces, the choice is straightforward. 
As a Janus particle is self-propelled along its orientation $\varphi$, the most natural fit function is the unitary vector $\ee_{\varphi} = (\cos(\varphi),\sin(\varphi))$, which provides the self-propulsion velocity $U$.
In contrast, the choice for the pair interactions is still broad.  
However, it is possible to further reduce the basis by setting additional constraints based on the symmetry of the problem. 

Assuming that the system is invariant under rotations and parity inversion, the interactions should be symmetric with respect to the axis defined by the particles 1 and 2:
\begin{align}
v_r(r,\theta_1,\theta_2)&=v_r(r,-\theta_1,-\theta_2),\\
v_\theta(r,\theta_1,\theta_2)&=-v_\theta(r,-\theta_1,-\theta_2), \\
\omega(r,\theta_1,\theta_2)&=-\omega(r,-\theta_1,-\theta_2),
\end{align}
We separate the angular and radial dependence by choosing a Fourier basis for the first one:
\begin{align}
v_r(r,\theta_1,\theta_2)&=\sum_{l=1}^{N_r} g_l(r)\sum_{n=0}^{N_o}\sum_{k=-n+1}^n a_{n,k,l}^r\cos([n-|k|]\theta_1+k\theta_2),\label{eq:coeffs_vr}\\
v_\theta(r,\theta_1,\theta_2)&=\sum_{l=1}^{N_r} g_l(r)\sum_{n=1}^{N_o}\sum_{k=-n+1}^n a_{n,k,l}^\theta\sin([n-|k|]\theta_1+k\theta_2),\label{eq:coeffs_vt}\\
\omega(r,\theta_1,\theta_2)&=\sum_{l=1}^{N_r} g_l(r)\sum_{n=1}^{N_o}\sum_{k=-n+1}^n a_{n,k,l}^\omega\sin([n-|k|]\theta_1+k\theta_2),\label{eq:coeffs_om}
\end{align}
and a smooth localized radial basis for the second:
\begin{equation}
g_l(r) = \frac{1}{l!} \left(\frac{r}{r_0} \right)^l\ed^{-r/r_0},
\end{equation}
whose maxima are given by $lr_0$ and the width by $r_0$. 
From the coefficients $a_{n,k,l}$, we also define the derived quantity $a_{n,k}(r)=\sum_l a_{n,k,l} g_l(r)$, the amplitude of the harmonic function indexed by $n,k$ as a function of the distance $r$.
Thus, the number of coefficients to determine is $N_b = N_r\times [1+3N_o(N_o+1)]$ with $N_r$ the number of radial functions and $N_o$ the order of the angular basis.
We emphasize that this decomposition is general and does not make any assumption on the form of the interactions, beyond the symmetry relations that they satisfy. 
 
Several comments can be made. 
First, there is a priori no reason to limit the distance $r$ at which pair interactions are inferred.
However, in practice, we set a maximum bound $r_\mathrm{max}$ to reduce the computation time. 
Similar to many particles simulations, the whole frame is divided into cells and the pair interactions are inferred only for a distance within the dimension of a cell.
This is not a problem as long as the inferred forces are negligible at the boundaries of the cell.

Second, although the coefficients $a_{n,k,l}$, or equivalently $a_{n,k}(r)$, represent in many cases the amplitude of harmonic terms contributing to the inferred velocity, their value may change with the order of the angular basis $N_o$.
This comes from the fitting procedure that adjusts the coefficients in order to minimize the error of the overall functional basis. 
If the chosen order $N_o$ is large enough to capture the dynamics, the underlying coefficients $a_{n,k}(r)$ could give, indeed, information on the harmonic composition of the velocities. 
On the contrary, their physical meaning as well as that of the inferred velocity are compromised if the dynamics is not captured properly. 
It is therefore important to assess the quality of the inference before discussing the meaning of the coefficients $a_{n,k,l}$ or $a_{n,k}(r)$. 

The last point relates to the diffusion inference. Here, we assume that the diffusion coefficients are constant with no cross diffusion. Moreover, the translational diffusion coefficient $D$ is too small to be measured with the experimental temporal resolution $\delta t$: the displacement due to self propulsion, $U\delta t$, is much larger than the one due to diffusion, $\sqrt{D\delta t}$, meaning that $\delta t\gg D/U^2$. As a consequence, it only reduces to the estimation of the rotational diffusion coefficient $D_r$.
Note that SFI can also deal with more complex situations such as state dependent diffusion~\cite{Frishman2020} for which the diffusion is expanded on a set of tensor functions. 


\section{Interactions in the flocking system}
\label{Sec:Flocking-System}

\begin{figure*}[t]
\begin{center}
\includegraphics[width=170mm]{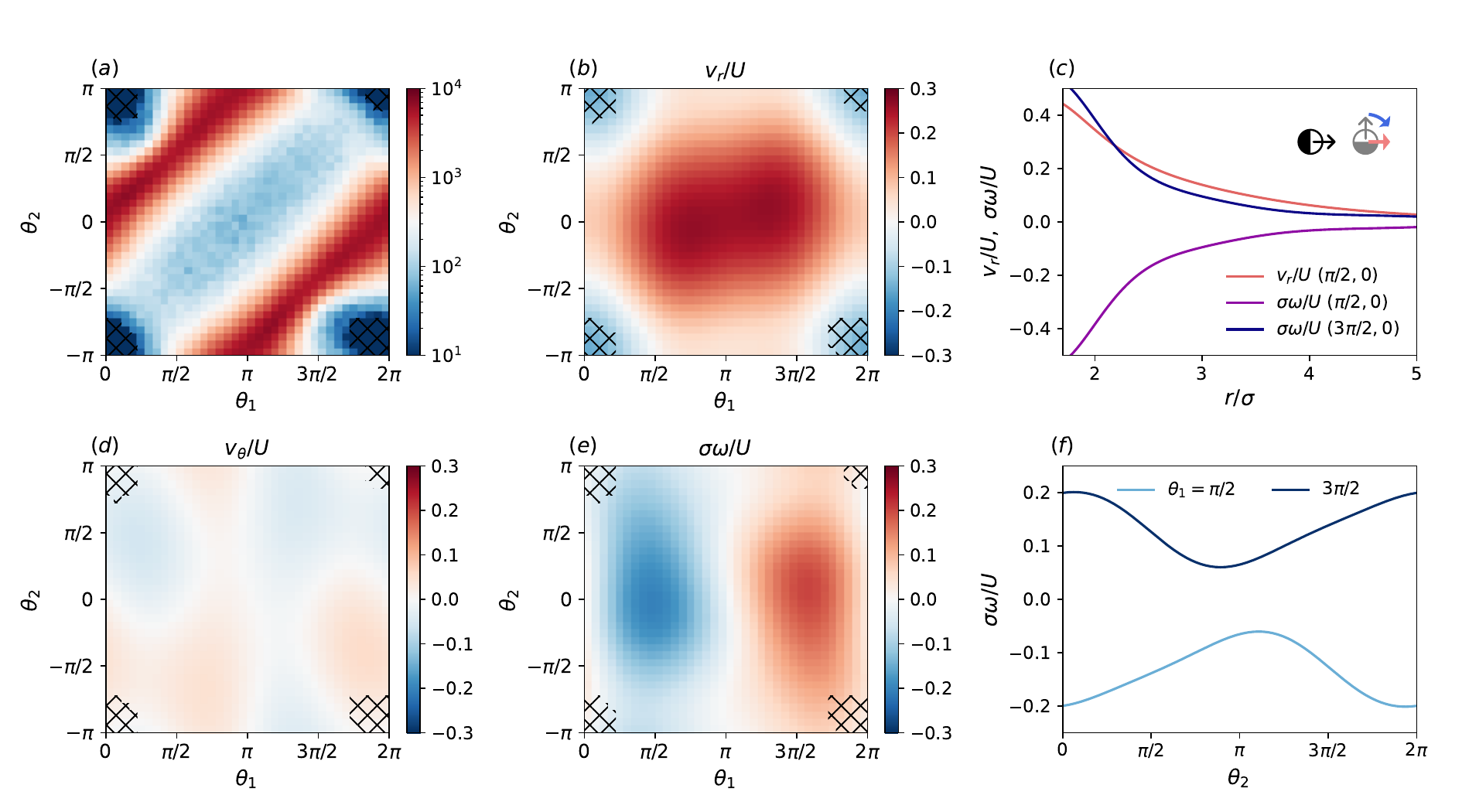}
\end{center}
\caption{SFI results for set $A$ (flocking system). (a) Pair correlation: histogram of the relative orientations at $r/\sigma=2.5$ with $\sigma= \qty{3.17}{\micro m}$ the particle diameter. (b,d,e) Inferred pair interactions at $r/\sigma=2.5$ as a function of the angles $\theta_1, \theta_2$: (b) radial, (d) azimuthal 
    and (e) angular velocities. The hatched areas correspond to areas of the histogram with lower statistics (bin count below $10$).
    (c) Radial and angular velocities as function of the distance $r$ for specific angular positions: $(\theta_1,\theta_2) = (\pi/2,0)$, $(3\pi/2,0)$. 
    Inset:  particle 1 (grey) and particle 2 (black) in the position $(\theta_1,\theta_2)=(\pi/2,0)$  with the red and blue arrows indicating the directions of the radial and angular velocities, respectively.
    (f) Angular velocity for fixed $\theta_1$ as a function of $\theta_2$. Detailed parameters of the inference are given in the Appendix.}
\label{fig:inferred_daiki}
\end{figure*}

\begin{figure*}[t]
\begin{center}
\includegraphics[width=170mm]{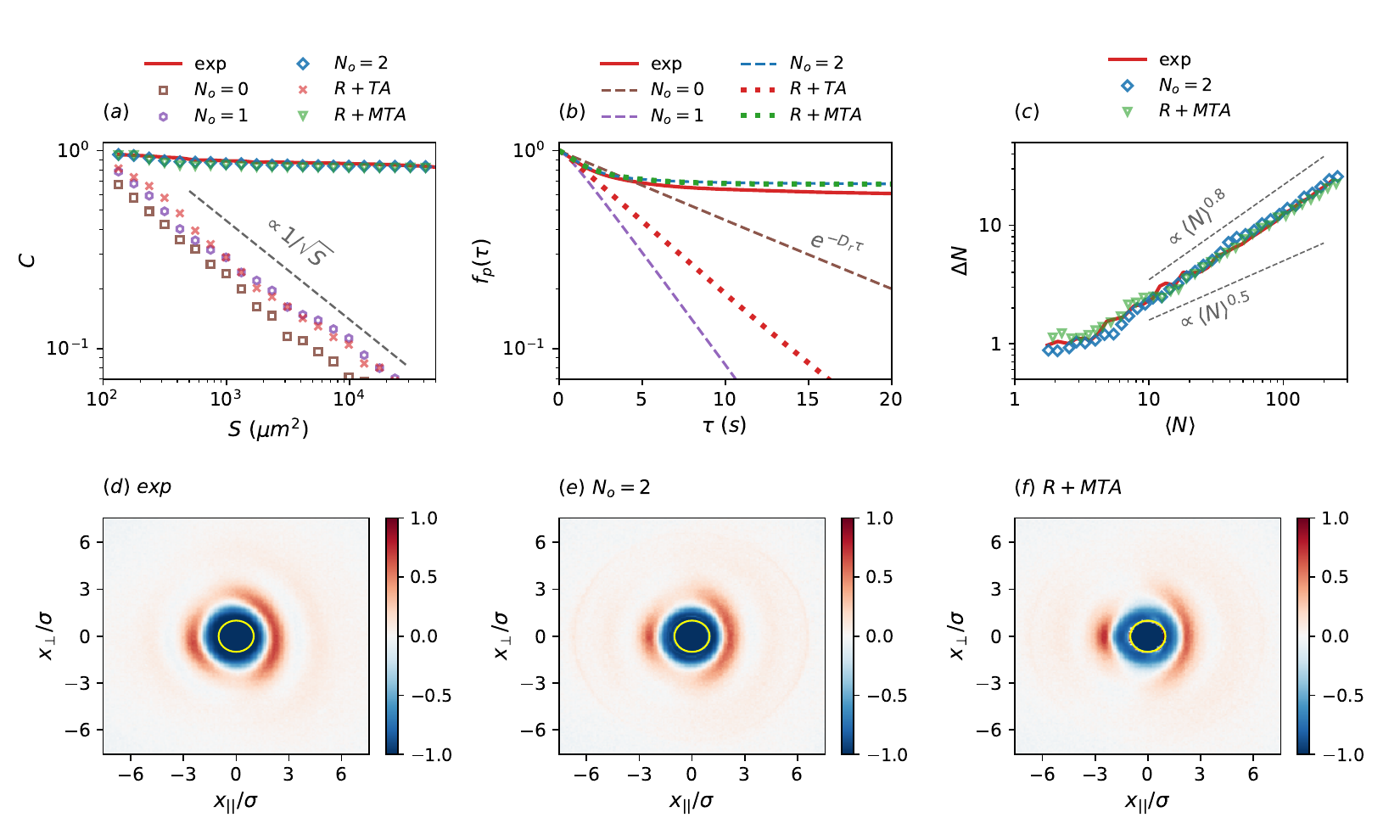}
\end{center}
\caption{Comparison of the observables between: the experimental flocking system (set $A$), the simulations with the inferred interactions at different angular orders $N_o=0,1,2$ and the simulations denoted R+TA, R+MTA including only certain terms at $N_o=2$. In all simulations, the frame size and the particle number are identical to the experiment. (a) Polar order parameter $C$ vs the area $S$ of the ROI. 
    (b) Polarity autocorrelation $f_p(\tau)$. (c) Number fluctuation $\Delta N$ vs average number of particle $N$. 
(d,e,f) Pair correlation function $B(\rr)$ respectively for (d) the experiment, (e) the simulation at $N_o=2$ and (f) the simulation R+MTA. The size of the particle is represented by the yellow circle in the center. See main text for the definition of the interactions in the simulations R+TA, R+MTA and the definition of the observables.}
\label{fig:sim_inferred_daiki}
\end{figure*}


\subsection{Results}

We turn to the experiments of Ref.~\cite{Iwasawa2021} and especially the data set $A$ for which the ordered phase is observed.
For this, we run SFI with the basis functions defined above at second order ($N_o=2$), which provides first the inferred self-propulsion speed $U\simeq\qty{9.7}{\micro m\per\second}$ and the rotational diffusion $D_r\simeq\qty{0.092} {\per\second}$.
These values are found in good agreement with those measured in Ref.~\cite{Iwasawa2021}. 

Then, we assess the reliability of the inference by analyzing the statistical distribution of the particles with respect to each other.  
Although our approach provides the interactions for all distances and orientations, the system does not explore uniformly all the configuration space, so that the interactions may not be inferred properly in unexplored regions. 
Indeed, the radial pair correlation (Fig.~\ref{fig:schematic}(d)) indicates the absence of neighbor particles at short distance $r/\sigma < 2$ with the particle diameter $\sigma=\qty{3.17}{\micro m}$, an accumulation around $r/\sigma = 2.5$ and a uniform presence beyond $r/\sigma > 3$. 
Since the distance $r/\sigma = 2.5$ is akin to a mean distance in the flock, we choose it and discuss the corresponding histogram of the relative orientations in the plane $(\theta_1,\theta_2)$ (Fig.~\ref{fig:inferred_daiki}(a)). 
Similar to the distance $r$, the distribution of neighbors is also not uniform. 
There is a region of high density, defined by $\theta_1 \simeq \theta_2$, which corresponds to particles moving in the same direction, revealing the flocking behavior. 
There is also a region of intermediate density and an unexplored one, located around $\theta_1\simeq 0$, $\theta_2\simeq \pi$, corresponding to the situation where the two particles are back to back. 
The presence or even the excess of particles consolidate the inference in these regions.
On the contrary, the inference in the unexplored region should be taken with care, although it can be qualitatively correct, for example in term of sign. 
For clarity, we hatch the questionable area if the bin count in the histogram falls below the arbitrary threshold of 10 occurrences in a pixel.

The inferred interactions $v_r$, $v_\theta$, $\omega$ exerted by a particle 2 on a neighbor particle 1 at a distance $r/\sigma = 2.5$ are presented in Fig.~\ref{fig:inferred_daiki}(b,d,e).
The interactions and distances are rescaled by $U$ and $\sigma$ to allow the comparison of the interaction components between each other and with the self-propulsion.
Note that in the plane $(\theta_1,\theta_2)$, a neighbor particle 1 is located in front of the particle 2 for $-\pi/2<\theta_2<\pi/2$ and behind it for $-\pi<\theta_2<-\pi/2$ or $\pi/2<\theta_2<\pi$.

The radial interaction $v_r$ is mainly repulsive for a large range of angular positions Fig.~\ref{fig:inferred_daiki}(b), so that the interaction prevents self-contact. 
In particular, there are two maxima located at $(\theta_1,\theta_2)\simeq (\pi/2,0)$ and $(\theta_1,\theta_2)\simeq (3\pi/2,0)$, that is, when particle 1 is located in front of the particle 2 and oriented sideways to it. 
Although the statistics suggest caution, it is probable that the interaction when particles are back to back is negligible or weakly negative. 
The azimuthal interaction $v_\theta$ has a non trivial shape but with an amplitude much smaller than the radial or the angular interactions (Fig.~\ref{fig:inferred_daiki}(d)).

In contrast, the amplitude of the torque (Fig.~\ref{fig:inferred_daiki}(e)) is, in dimensionless units, as strong as the radial interaction, but with a more complex dependence on the orientations.  
It is negative for $0\leq \theta_1\leq \pi$ and positive for $\pi\leq \theta_1\leq 2\pi$, which means that the particle 1 always turns away from the particle 2.
Such ``turning away'' interaction, in addition to an isotropic radial repulsion, has been identified as being responsible for the flocking in Ref.~\cite{Das2024Flocking}.
However, it should be noted that, here, the interaction is strongly modulated by the location of the neighbor particle 1 relative to particle 2: it is maximal when the particle 2 faces the particle 1 ($\theta_2\simeq 0$) and almost vanishes when the particle 2 turns its back on particle 1 ($\theta_2\simeq \pi$). 
Another representation is proposed in Fig.~\ref{fig:inferred_daiki}(f), where we see that the amplitude of the torque can significantly change with $\theta_2$ for fixed $\theta_1$. 
Due to its strong magnitude, we hypothesize that it could play an important role in the flocking.
We come back to this question in the next section.

Finally, Fig.~\ref{fig:inferred_daiki}(c) presents the evolution of the radial and angular interactions as a function of the distance $r$, for an orientation $\theta_1$, $\theta_2$ that maximize the interactions. 
We observe that in both cases the interactions decrease monotonically and vanish above $r/\sigma \gtrsim 5$.
These behaviors of the interactions are robust features supported by sufficient statistics.

\begin{figure*}[t]
\begin{center}
\includegraphics[width=170mm]{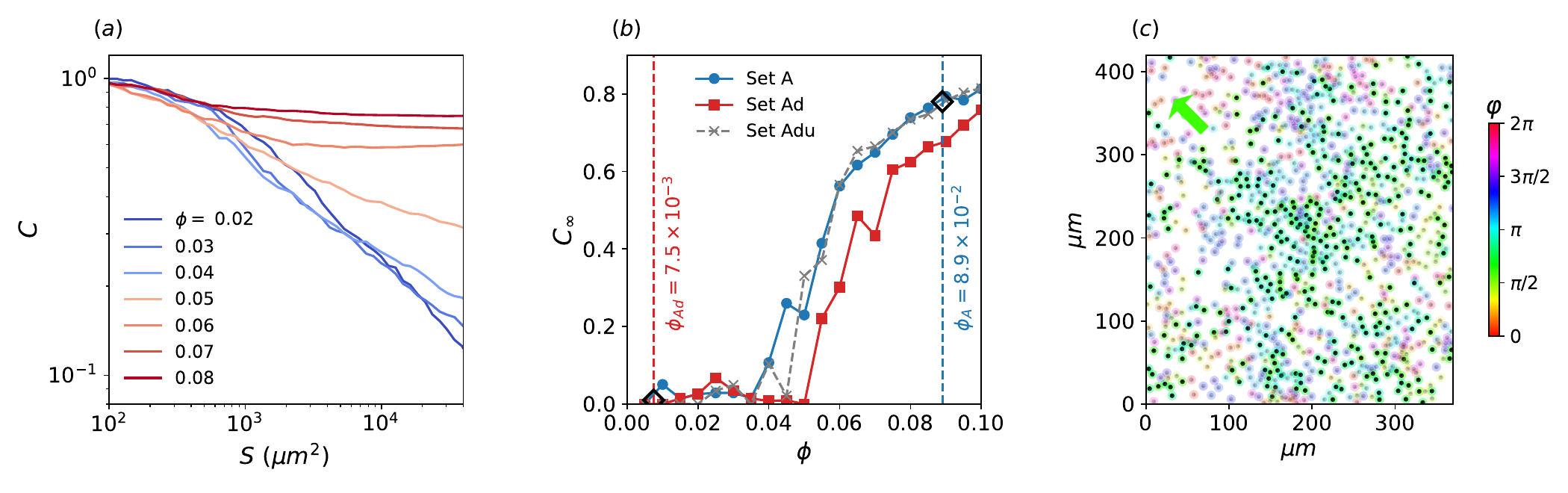}
\end{center}
\caption{Inferred simulations with varying area fraction $\phi$. (a) Evolution of the function $C(S)$ with $\phi$ on a log-log scale, with the inferred parameters extracted from set $A$. (b) Evolution of the parameter $\Cinf$ with $\phi$, with the inferred parameters of set $A$, $Ad$ and $Adu$. The value $\Cinf$ fits the relation : $C=C_{\infty}+kS^{-\gamma/2}$. The parameters $Adu$ are the same as for $Ad$ except the value $U$ equal to that of set $A$. The vertical dashed lines indicate the area fractions $\phi_A$, $\phi_{Ad}$ of the experimental sets $A$ and $Ad$, the black diamonds indicate the related experimental values $\Cinf$. (c) Snapshot of the data set taken close to the transition ($\phi=\num{5e-2}$), with the inferred parameters of set $A$.}   
\label{fig:flocking}
\end{figure*}

\begin{figure*}[t]
     \centering
    \includegraphics[width=170mm]{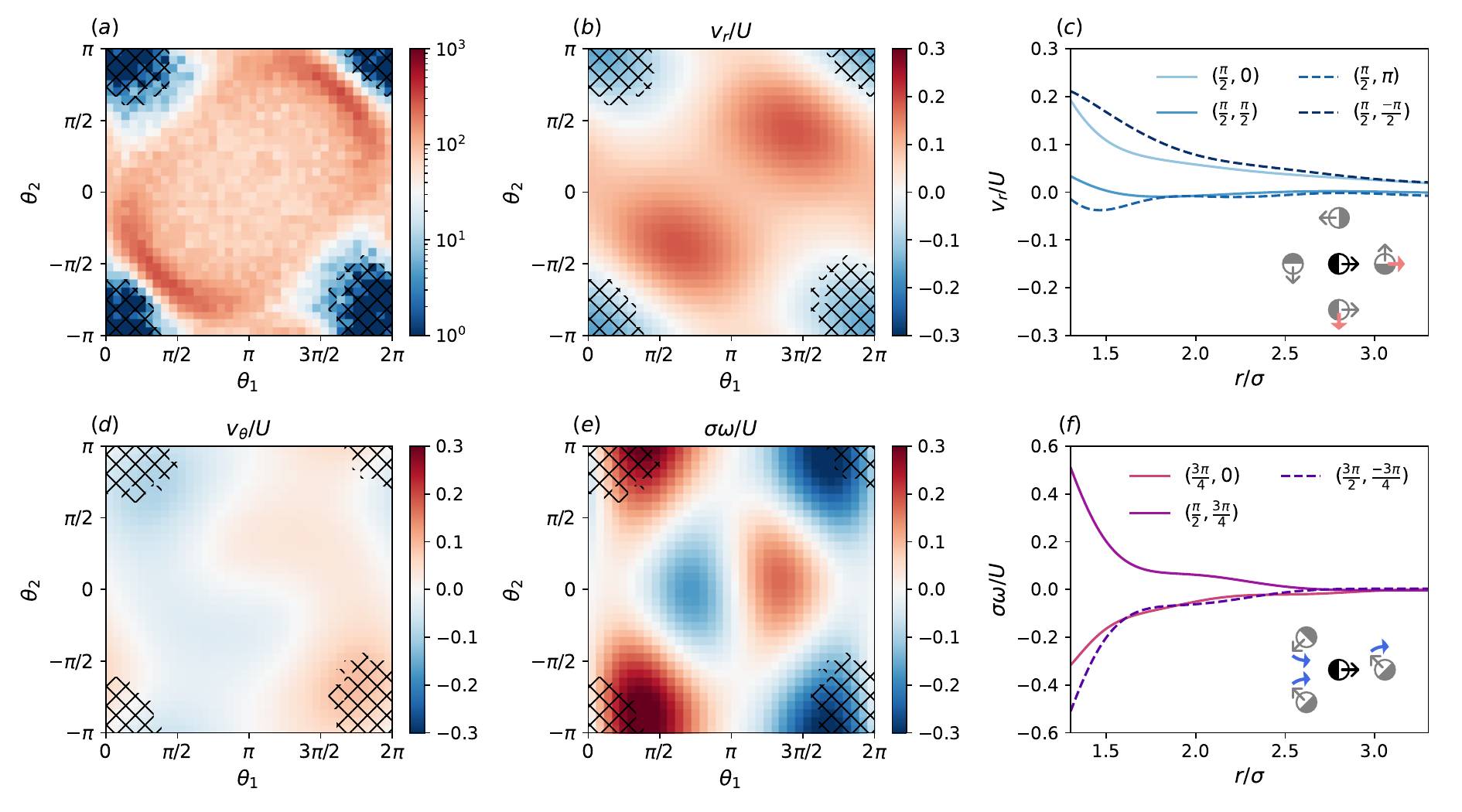}
    \caption{SFI results for set $B$ (disordered system). (a) Histogram of the relative orientations at $r/\sigma=1.5$ with $\sigma=3.17\ um$ the particle diameter. (b,d,e) Inferred pair interaction at $r/\sigma=1.5$ as a function of the angles $\theta_1$, $\theta_2$. (b) radial, (d) azimuthal and (e) angular velocities. As before, the areas with lower statistics are hatched (bin count below $10$). 
    (c) Radial velocity as function of the distance $r$ for specific angular positions $(\theta_1,\theta_2)$; inset: particle 2 (black) and particle 1 (grey) at different positions, with the red arrows indicating the direction of the radial velocity.
    (f) Angular velocity as function of the distance $r$ for specific angular positions; inset: particle 2 (black) and particle 1 (grey) at different positions, with the blue arrows indicating the direction of the angular velocity.
    }  
    \label{fig:inferred_alexis}
\end{figure*}

\subsection{Numerical simulations} 

To validate the inferred interactions, we run numerical simulations with these interactions and compare the observables with those measured in experiment. 
The results are presented in Fig.~\ref{fig:sim_inferred_daiki} for different orders $N_o=0,1,2$ of the angular basis. 
Overall, we find that the synthetic ordered phase, obtained at the second order, presents the same statistical properties than the experimental one, and that this order is also the minimal one required to describe faithfully the dynamics.

For example, Fig.~\ref{fig:sim_inferred_daiki}(a) shows the polar order parameter $C=\langle | \langle e^{i\phi} \rangle_S| \rangle_t$ for different sizes $S$ of a region of interest (ROI), averaged over temporally uncorrelated frames. 
The order parameter, for both the experiment and the simulation, levels off towards the same finite value $\Cinf\simeq0.8$ for $S>10^4\unit{\micro m^2}$, which indicates the existence of a long range order. 
Fig.~\ref{fig:sim_inferred_daiki}(b) shows the evolution of the polarity autocorrelation function $f_p(\tau)=\langle \ee_\varphi(t)\cdot \ee_\varphi(t+\tau)\rangle_t$ where $\ee_\varphi(t)$ stands for the polarity of the particle at time $t$. 
Again, a plateau is observed in both cases, indicating a high level of correlation maintained even after a long period $\tau>\qty{10}{\second}$.
The same behavior is also retrieved for the number fluctuations $\Delta N$ with the scaling~\cite{Marchetti2013, Nishiguchi2023Deciphering} $\Delta N \sim \langle N \rangle^{0.8}$, Fig.~\ref{fig:sim_inferred_daiki}(c) and also for the pair correlation $B(\rr)$ integrated over the angle of particle 2,  Fig.~\ref{fig:sim_inferred_daiki}(d,e). 
$B(\rr)$ indicates the variation of the density of particles at a position $\rr$ with respect to a particle moving in the direction $\ee_\varphi$.
It is constructed from the polar pair correlation $C(\rr,\theta,\theta')$, which from rotational invariance is contained in $C(\rr,0,\theta')$: $B(\rr) = (1/2\pi) \int_{0}^{2\pi} C(\rr,0,\theta')d\theta'$.
In particular, note in both cases the accumulation of particles around $r/\sigma=2.5$ (the red ring in Fig.~\ref{fig:sim_inferred_daiki}(d,e)).

In comparison, the interactions inferred at zero or first order lead to incorrect predictions. 
The particles do not flock and the system remains disordered. 
This is evidenced in Fig.~\ref{fig:sim_inferred_daiki}(a) where the polar order function $C(S)$ decreases as $\propto 1/\sqrt{S}$, which is characteristic of uncorrelated movements or in Fig.~\ref{fig:sim_inferred_daiki}(b) where the the polarity function $f_p(\tau)$ decreases as $e^{-D'\tau}$ with the effective diffusion constant $D'$.
Note that, at zeroth order, the constant $D'$ is identical to the rotation diffusion $D_r$, whereas at first order $D'$ increases, so the diffusion is enhanced in this case.
The pair correlation function $B(\rr)$ at zeroth or first order is also very different compared to the one observed in the experiment (not shown). 

Besides, we find instructive to search for the dominant terms that can explain the main features of the experimental dynamics.
From the inferred interactions at second order, Fig.~\ref{fig:inferred_daiki}(b,d,e), it directly appears that some terms can be neglected, such as the azimuthal interaction $v_{\theta}$ due to its low amplitude. 
For the radial and angular interactions, the choice of the important terms is less obvious.  
However, we can identify them by turning on and off certain terms of the interaction in the simulations and see the effect on the dynamics. 
First, we choose to keep only the isotropic radial repulsion $a^r_{0,0}$ and the $\theta_2$-independent ''turning away'' interaction $a^\omega_{1,0}$ (denoted R+TA in Fig.~\ref{fig:sim_inferred_daiki}).
Thus, the angular dependence of the interaction is identical to the one studied in Ref.~\cite{Das2024Flocking}.
We see that the system remains disordered (Fig.~\ref{fig:sim_inferred_daiki}(a,b)).  
Second, in contrast, we choose to keep both the radial repulsion $a^r_{0,0}$ and the full form of the angular interaction, which thus includes the modulation as a function of $\theta_2$ (denoted R+MTA, for ``Modulated Turning Away'' in Fig.~\ref{fig:sim_inferred_daiki}).
With these interactions, the flocking is recovered, as well as the various observables of the experimental system. 
We conclude that the full form of the angular interaction, in addition to an isotropic radial repulsion, is important and contributes to the flocking behavior of the experimental system. 

\subsection{Extrapolation to different densities} 

Another advantage of inferring the pair interactions between the particles is to be able to predict a behavioural change in the system as a function of parameters such as the particle density. 
To illustrate this possibility, we run numerical simulations with the inferred parameters obtained previously ($U$, $D_r$ and the pair interactions at second order for set $A$), and then vary the area fraction from a low density $\phi=\num{5e-3}$ to a high density $\phi=0.1$.
The behavioural change is quantified by the evolution of the polar order function $C(S)$ and the parameter $C_{\infty}$, extracted from the relation $C=C_{\infty}+kS^{-\gamma/2}$, Fig.~\ref{fig:flocking}. 
We observe two regimes. 
For small area fractions, $\phi < \num{4e-2}$, the system is disordered, $C\propto S^{-1/2}$ and $C_{\infty}\simeq0$.
In contrast, for higher values $\phi > \num{6e-2}$, we retrieve the flocking behavior observed in the experimental set $A$, Fig.\ref{fig:schematic}(c), $C$ converging to $ C_{\infty} \simeq 0.8  $ for $ S > 10^3\ \unit{\micro m^2} $. 
At the threshold $ \num{4e-2} < \phi < \num{6e-2}$, the two phases coexist.
For example, Fig.~\ref{fig:flocking}(c) shows the emergence of a swarm of particles moving along the same direction for $\phi=\num{5e-2}$. 

The existence of the two dynamical regimes can be verified experimentally: Ref.~\cite{Iwasawa2021} reports another data set at lower area fraction $\phi_{Ad}=\num{7.5e-3}$, which we denote set $Ad$, where the system remains disordered.
Since only the density is varied between the two data sets, one should expect to find identical pair interactions in both cases, although an increase of the self-propulsion velocity ($U=\qty{14.1}{\micro m\per s}$) compared to the ordered phase ($U=\qty{9.7}{\micro m\per s}$) is reported. 
We run SFI for the set $Ad$ with the same parameters as before, and run simulations with the inferred interactions for different area fractions $\phi$. 
We see in Fig.~\ref{fig:flocking}(b) that there is almost no difference between sets $A$ and $Ad$:
the two dynamical regimes are retrieved, indicating that the inference is not altered by the particle density or the state of the system.
The shift between the two curves is due to the change in $U$. 
Indeed if $U$ in the simulations with the interactions inferred from the set $Ad$ is taken identical to the $U$ obtained in set $A$, then the two curves become identical (compare sets $A$ and $Adu$), so that the pair interactions are indeed identical in both cases.
The variation of the self-propulsion speed between sets $A$ and $Ad$ could come from variations in the distance between the electrodes, resulting in different strengths of electric field.
The inference applied to a reduced set of particles or applied to a single data set at given area fraction, can help to predict the behavior at any other density and to reveal possible phase transitions, which could be useful to guide experiments. 

\section{Interactions in the disordered system}
\label{Sec:Disordered-System}

\begin{figure*}[h]
    \centering
    \includegraphics[width=160mm]{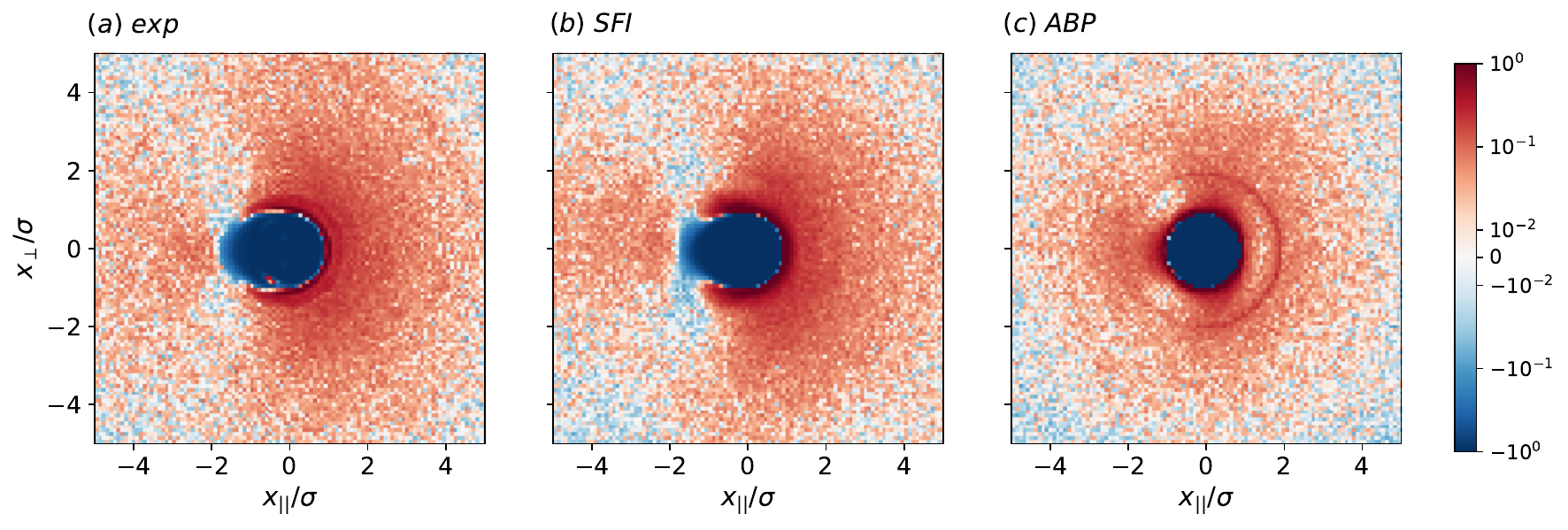}
    \caption{Evaluation of the pair correlation function $B(\bf{r})$ for set B for a similar level of statistics (frame number $\num{8e3}$, particle number $\sim 500$). (a) Experimental measure~\cite{Poncet2021}. (b) Simulated data with the inferred interactions. (c) Simulated data for Active Brownian Particles (ABP).}  
    \label{fig:pcorel_alexis}
\end{figure*}


\subsection{Results}

We turn now to set $B$ provided by Ref.~\cite{Poncet2021}, where the system remains in a disordered state. 
One goal of this study was to provide a pure experimental ABP system, that is with only radial pair interactions, in order to compare with the theoretical predictions.
As a consequence, the experimental parameters (electrical field, concentration of sodium chloride) were tuned so as to decrease the overall hydrodynamic and electrostatic interactions. 
Indeed, we see the particles moving in a disordered way with no polar order measured, and thanks to the weak interactions, the particles collide with one another. 
The dynamics looks qualitatively very similar to the one expected for pure Active Browian Particles (ABP).
However, if one takes a finer observable, such as the pair correlation $B(\rr)$, the experimental one shows the formation of a depletion zone just behind the particle which is absent for ABP.
It is thus possible that residual interactions are still present in the system, which could be characterized. 

In contrast to the first dataset, the detection of the force is more challenging, because the interactions are much weaker and take place at smaller distances between particles, within $1<r/\sigma<3$. 
At such distance, the inference may be blurred by the collisions, especially if the time step is a bit long, as is the case here where $U\delta t/\sigma\simeq0.2$. 
We also note that the error on the orientation detection is higher $\varphi_\mathrm{err}\approx\qty{9}{\degree}$, which also decreases the measurement precision. 
Despite these difficulties, we shall see that it is possible to obtain the interactions in this system. 

Once again, we run SFI with the basis functions defined at second order ($N_o=2$). 
The inference provides the active velocity $U\simeq \qty{7.0}{\micro m\per\second}$ and the rotational diffusion $D_r\simeq \qty{0.085}{\per\second}$, which are similar to those reported in Ref.~\cite{Poncet2021}.
Then, we choose the distance $r/\sigma=1.5$ and discuss the histogram of the relative orientations presented in Fig.~\ref{fig:inferred_alexis}(a). 
Since the system is disordered, the distribution of neighbors is quite homogeneous along $\theta_1$, $\theta_2$ (Fig.~\ref{fig:inferred_alexis}(a)). 
The greater densities are observed around $(\theta_1,\theta_2)=(\pi/2,-\pi/2)$ or $(3\pi/2,\pi/2)$. 
Here, the unexplored region where the particles are back to back appears larger than compared to set $A$, because of the closer distance $r/\sigma=1.5$. 
Again, the questionable areas are hatched and reported on top of the inferred interactions.

The inferred interactions $v_r$, $v_\theta$, $\omega$ at distance $r/\sigma=1.5$ are presented in Fig.~\ref{fig:inferred_alexis}(b,d,e).
First, the radial interaction Fig.~\ref{fig:inferred_alexis}(b) looks similar to the previous experiment. 
It is still positive when the two particles face each other, and is maximally repulsive for two angular positions : $(\theta_1,\theta_2)\simeq (2\pi/3,-\pi/3)$ or $(4\pi/3,\pi/3)$, that is, when the particles fly by each other.  
In contrast, the interaction is weakly attractive when the particles are back to back $(\theta_1,\theta_2)\simeq (0,\pi)$, although this last region should be taken with care because of the poor statistics, as indicated by the hatched area.
As in the previous experiment, the azimuthal interaction is very weak (Fig.~\ref{fig:inferred_alexis}(d)).

The most surprising feature is the presence of a relatively strong angular interaction, which was not considered in Ref.~\cite{Poncet2021} (Fig.~\ref{fig:inferred_alexis}(e)).
The particle 1 turns away from particle 2 when particle 2 is oriented towards particle 1 ($\theta_2\simeq 0$) and it turns towards particle 2 when particle 2 turns its back on particle 1 ($\theta_2\simeq \pi$).
Figs.~\ref{fig:inferred_alexis}(c,f) show the evolution of the radial and angular velocities with the distance $r$ for specific positions. In both cases, the interactions monotonically decrease and vanish above $r/\sigma\simeq3$. 

\subsection{Numerical simulations} 

To validate the inference, we run numerical simulations with the inferred interactions and focus on the pair correlation function $B(\rr)$. 
The experimental result is recalled in Fig.~\ref{fig:pcorel_alexis}(a).
We observe an accumulation zone in front of the particle, and a depletion zone in the back. Two depletion wings emerge in the back of the particle.  
To compare, we plot in Fig.~\ref{fig:pcorel_alexis}(b) the pair correlation function of the simulation with the inferred interactions.  
All the features described previously are retrieved: the accumulation and depletion zones, as well as the depletion wings. 
In contrast, Fig.~\ref{fig:pcorel_alexis}(c) shows the pair correlation for ABP with a soft sphere potential interaction. 
In this case, we see two accumulations zones in front and back of the particle with two weak depletion wings emerging on the sides of the particle~\cite{Poncet2021}, which is significantly different than in the previous figures.

Therefore, we see that despite the overall decrease of the interactions, the ABP assumption remains too crude to explain all the features of the dynamics. 
The pair correlation function $B(\rr)$ is correctly recovered if one takes into account also the angular interactions present at short distance around the particle.
Furthermore, this example is interesting from the viewpoint of the method, as it shows that, in spite of the multiple sources of noise, SFI is sensitive enough to capture weak interactions.

\section{Symmetry considerations}
\label{Sec:Origin}

SFI allowed us to determine precisely the interactions between the particles, but it does not give any information about their origin.
As outlined in the introduction, none of the electrostatic or hydrodynamic interactions can be determined analytically.
However, these interactions have specific properties, which can help to discriminate them. 
For hydrodynamic interactions, the effect of the particle 2 on the particle 1 can be expected to be dominated by the advection of the particle 1 by the flow created by the particle 2. 
As such, it should not depend on the orientation of the particle 1, so that $\theta_1$-dependent terms can mainly be attributed to electrostatics.
Since the interactions are strongly dependent on $\theta_1$, we conclude that the electrostatic interaction is at least present if not dominant in both sets $A$ and $B$. 

In turn, the electrostatic interaction imposes that the forces derive from an effective potential $V(r,\theta_1,\theta_2)$~\cite{Yan2016Reconfiguring}:
\begin{align}
v_r & = -\mu \frac{\partial V}{\partial r},\label{eq:pot_vr}\\
v_\theta & = \frac{\mu}{r} \left[\frac{\partial V}{\partial \theta_1}+\frac{\partial V}{\partial \theta_2} \right],\label{eq:pot_vt}\\
\omega & = -\mu_r \frac{\partial V}{\partial \theta_1},\label{eq:pot_om}
\end{align}
where $\mu$ and $\mu_r$ and the translational and rotational mobilities.
As a consequence, the interactions should follow the relations: 
\begin{align}
  \frac{\partial v_r}{\partial\theta_1} &\propto \frac{\partial\omega}{\partial r}, \label{eq:sym_vr_om} \\
  \frac{\partial (rv_\theta)}{\partial r} & = - \frac{\partial v_r}{\partial\theta_1} - \frac{\partial v_r}{\partial\theta_2},
\end{align}
where the proportionality involves the unknown ratio of translational and rotational mobilities.
Since $v_r(r,\theta_1,\theta_2)$, $\omega(r,\theta_1,\theta_2)$ are measured with a good signal to noise ratio in sets $A,B$, Eq.~(\ref{eq:sym_vr_om}) becomes an interesting criteria to test the presence of the electrostatic interaction in those systems. 


\begin{figure*}[t]
    \centering
    \includegraphics[width=170mm]{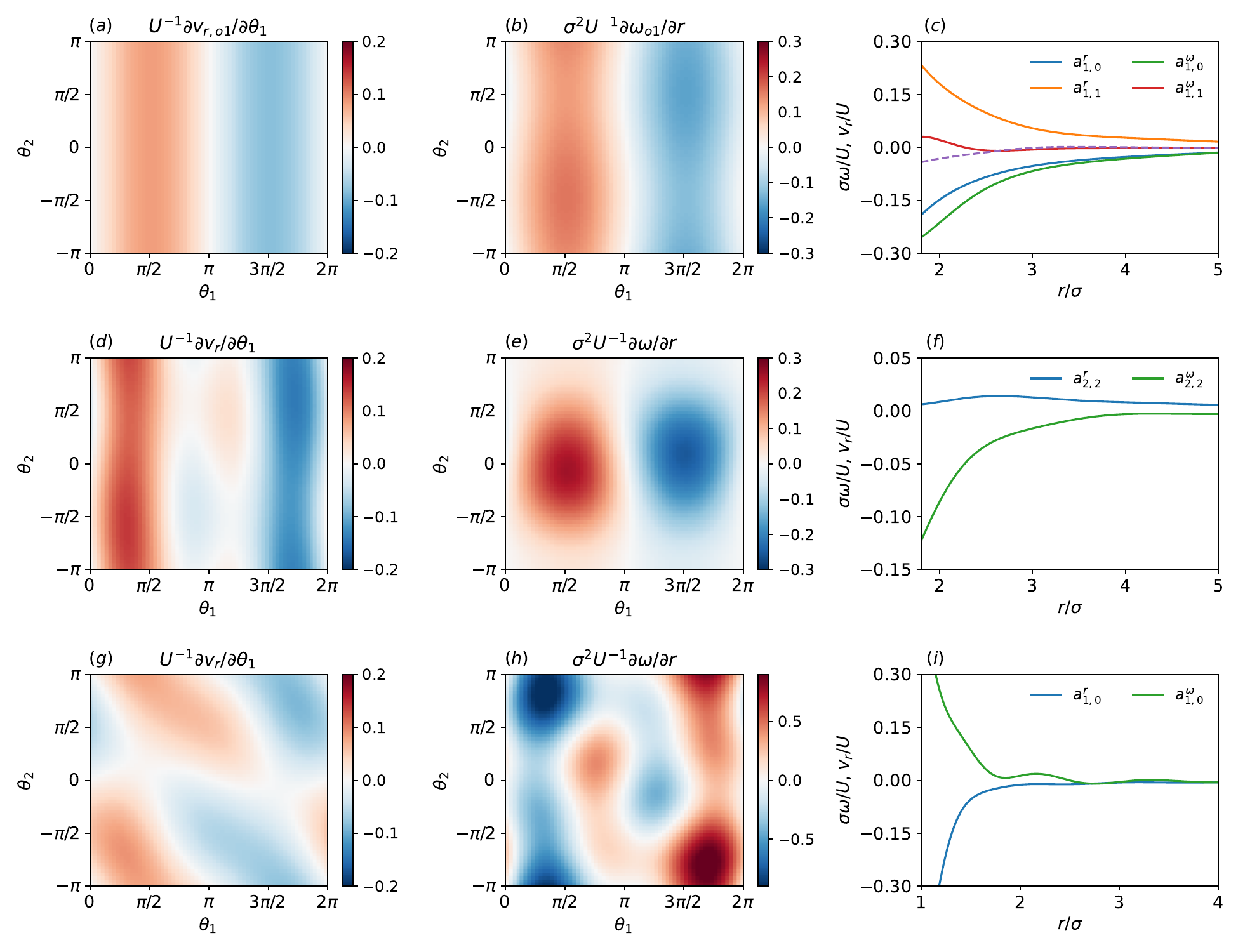}
    \caption{Evaluation of the symmetry relation Eq. \ref{eq:sym_vr_om}: (a,b) For set A including only the first order term of the angular basis at $r/\sigma=2.5$, (c) evolution of the coefficients $a_{n,k}(r)$ with the distance. The dashed line represents the difference $|a_{1,0}^{r}|-|a_{1,1}^{r}|$. (d,e) For set A including all the terms up to the second order at $r/\sigma=2.5$, (f) evolution of $a_{2,2}^{r}(r)$, $a_{2,2}^{\omega}(r)$ with the distance $r$. (g,h) For set B including all the terms up to the second order at $r/\sigma=1.5$, (i) evolution of $a_{1,0}^{r}(r)$, $a_{1,0}^{\omega}(r)$ with the distance. The function $a_{n,k}(r)$ stands for the coefficient of the harmonic term $\cos((n-|k|)\theta_1+k\theta_2)$ or $\sin((n-|k|)\theta_1+k\theta_2)$, see notations in Eq. \ref{eq:coeffs_vr}, \ref{eq:coeffs_vt}, \ref{eq:coeffs_om}. 
    \label{fig:symmetries}}
\end{figure*}

Fig. \ref{fig:symmetries}(a,b) shows the relation (\ref{eq:sym_vr_om}) in case of set $A$ and for only the first order terms of the angular basis. 
Overall, we see a good agreement which implies a specific behavior for each harmonic terms, see Fig.~\ref{fig:symmetries}(c). 
For example, the term $a_{1,1}^{\omega}$ is almost zero as expected since $\partial \omega/\partial\theta_2\simeq0$ (Fig.~\ref{fig:inferred_daiki}(e)). 
The terms $a_{1,0}^{\omega}$ and $a_{1,0}^{r}$ are of same sign, which can be explained as follows. 
Suppose $a_{1,0}^{\omega}\propto r^{-n}$, then $a_{1,0}^{r}\propto n/r^{n+1}$ and the ratio is necessarily $a_{1,0}^{\omega}/a_{1,0}^{r}>0$.
In particular, it justifies the electrostatic origin of the ``turning away'' interaction observed for the angular velocity since the amplitude $a_{1,0}^{\omega}$ has the expected sign. 
Besides, in case of electrostatic interactions, the potential should be invariant under reflection with respect to the axis defined by the particles $1$ and $2$, $V(r,\theta_1,\theta_2) = V(r,-\theta_1,-\theta_2)$, but also and under the exchange of the particles, $V(r,\theta_1,\theta_2) = V(r,\theta_2+\pi,\theta_1+\pi)$.
This additional symmetry implies the equality $a_{1,0}^{r}=-a_{1,1}^{r}$ which can be verified as $|a_{1,0}^{r}|-|a_{1,1}^{r}| \simeq 0$ (dashed line in Fig.~\ref{fig:symmetries}(c)).
Therefore, we find at first order all the hallmarks of a dominant electrostatic interaction. 

However, Figs. \ref{fig:symmetries}(d,e,f) show that the symmetry relations do not hold anymore if the second order terms are included.
For example, the terms $a_{2,2}^{r}$ and $a_{2,2} ^{\omega}$ behave very differently with the distance $r$, which is not compatible with the existence of an effective potential.  
For set $A$, we must conclude that there is another component present in the system, probably of hydrodynamic type, which does not derive from a potential. 

In a similar way, we conclude that the interactions in set $B$ are a mix of electrostatic and hydrodynamic as the symmetry relations are also not satisfied (Figs. \ref{fig:symmetries}(g,h)).
Note that in this case, the violation is even stronger since it is not verified at first order; for instance the terms $a_{1,0}^{r}$ and $a_{1,0}^{\omega}$ are of opposite sign (Fig. \ref{fig:symmetries}(i)). 
In set $B$, it is thus likely that the electrostatic and hydrodynamic interactions are of same order of magnitude. 

\section*{Conclusions}

We have used Stochastic Force Inference (SFI) to learn the pair interactions between self-propelled Janus colloids from experimental trajectories.
In particular, we have studied two specific data sets: a dense set of flocking particles and another one where the system remains disordered. 
Although in both cases the experimental conditions were not specifically optimized for this inference task, we successfully measured the pair interactions, as evidenced by the ability to reproduce all the statistical properties of the experimental systems in numerical simulations. 
Moreover, the precise determination of the interactions allowed us to gain information about the nature of the interactions. 
Based on the symmetry relations that an electrostatic potential should satisfy, we find that in the flocking system the interactions are mainly electrostatic, with the presence of an additional hydrodynamic component that contributes sensitively to the dynamics. 
In contrast, the relative contributions between the electrostatic and hydrodynamic interactions are more balanced in the second experiment. 

Such information can be crucial, especially for the study of Janus colloids propelled by AC electric field. 
The behavior of these systems under different experimental parameters can be predicted  ~\cite{Yan2016Reconfiguring, Zhang2017, Zhang2021}, but these theoretical predictions have been obtained by focusing on electrostatics only and neglecting the other influences, such as the hydrodynamic flow.
Our results confirm the global shape of the interactions but also unveil unpredicted and quantitatively significant dependencies.
To further characterize the interactions in these systems, one could infer them in experiments that use different parameters such as the amplitude and the frequency of the electric field.

Applied to experiments where the interactions are unknown, such as Janus colloids self-propelled through phoretic effects~\cite{Ginot2015}, the inference of interactions could be a decisive step towards understanding and modeling the system.
Whether a preliminary qualitative knowledge of the interactions exists or not, their precise determination may allow one to use numerical simulations to explore the effect of changing parameters such as the density or the system size.

An important difficulty that arises when learning the interactions is the orientation detection, albeit the orientation is clearly visible on experimental images. 
In other cases, such as rolling colloids~\cite{Bricard2013}, the orientation is not even visible on the images, preventing the application of the current framework.
To infer the interactions in such case, a possible route is to treat the hidden orientation of the particles as a memory, and to extend SFI to use multiple time steps to infer the interactions.
This extension would allow to deal with systems where memory is embedded in the surrounding fluid, such as for light-activated colloidal microswimmers in a viscoelastic fluid~\cite{Narinder2018} or self-propelling droplets that modify their environment by leaving chemical footprints~\cite{Hokmabad2022Chemotactic}.

\section*{Appendix}
\subsection*{Experimental parameters.}

The experimental setup for both set $A$ and $B$ is briefly described here; more information can be found in the original articles~\cite{Iwasawa2021,Poncet2021}.
In the two experiments, the particles under study are silica based colloids with a diameter of $\sigma = \qty{3.17}{\micro m}$, whose half hemisphere is coated with a titanium layer. 
The particles are suspended in a NaCl solution, that is sandwiched between two ITO electrodes with a spacer of thickness $h$. 
Then, a perpendicular AC electric field of frequency $f$ and amplitude $E$ is applied, which ensure the self-propulsion of the particles. 

For the video recording, a CMOS camera (Baumer LXG-80 $3000\times2400$ pixels, 8 bit grayscale) is mounted on an inverted microscope (Olympus IX70 or Nikon EXLIPSE TE2000-U). 
To increase the contrast of the hemispheres, a green filter is inserted between the sample and the halogen lamp.
Hence, in the images, the metallic hemisphere appears in dark gray and the dielectric side in light gray. 
In addition, several routines based on Python libraries are used for the tracking of the particles.
In particular, the position of the particle is provided by the Hough transformation whereas the orientation is given by the center of mass of the pixels. 
These allows to extract for each particle, its position $\rr_i(t) = (x_i(t),y_i(t))$ and its orientation $\varphi_i(t)$ over time.

\textit{Parameters for set $A$~\cite{Iwasawa2021}}. 
Video clip of $3.1\times10^3$ frames with time interval $\delta t= \qty{0.067}{\second}$, area fraction $\phi_A=\num{9.7e-2}$ with around $\num{1.7e3}$ particles in a frame of size $420\times\qty{336}{\micro m^2}$. 
The particles move with the metallic (dark) face ahead in deionized water with an electrode separation $h=\qty{130}{\micro m}$ and an applied AC electric field of frequency $f=\qty{1}{\mega\hertz}$ and amplitude $E=\qty{1.23e5}{\Vpp \per m}$. 
Standard error for the detection of the orientation: $\varphi_\textrm{err}\simeq\qty{2}{\degree}$. 

\textit{Parameters for set $Ad$~\cite{Iwasawa2021}}. Video clip of $\num{3.1e3}$ frames with time interval $\delta t = \qty{.1}{s}$, area fraction $\phi_{Ad}=\num{7.5e-3}$, $\sim 400$ particles in a frame of size $ 735\times \qty{588}{\micro m^2}$. 
Same experimental condition as for set $A$. 
Standard error for the orientation: $\varphi_\textrm{err}\simeq\qty{12}{\degree}$.

\textit{Parameters for set $B$~\cite{Poncet2021}}. Video clip of $\num{8.4e3}$ frames with time interval $\delta t=\qty{.1}{\second}$, area fraction $\phi_B=\num{4e-2}$ with around $\num{5e2}$ particles in a frame of size $360\times\qty{288}{\micro m^2}$. Here, the particles move with the dielectric (white) face ahead for $h=\qty{50}{\micro m}$ in a sodium chloride solution of concentration $\qty{1e-4}{\mole\per\liter}$ and an applied AC electric field of frequency $f=\qty{5}{\kilo\hertz}$ and amplitude $E=\qty{2e5}{Vpp \per m}$.
Standard error for the orientation: $\varphi_\textrm{err}\simeq\qty{9}{\degree}$. 

\subsection*{SFI parameters.}


For both experiments, the Stratonovitch and Vestergaard estimators have been used for the velocities (also called ``drift'') and for the rotational diffusion coefficient, respectively.
The rectangular approximation has been used for the integrals.

\textit{Set $A$, Fig~\ref{fig:inferred_daiki}}. Parameters for the basis: $N_o=2$, $N_r=8$ and $N_b=152$. The index $l$ for the radial functions $g_l(r)$ spans $l \in \{ 0,2,4,\dots14 \}$ with $r_0=\qty{1}{\micro m}$. Cell dimension for the SFI algorithm: $r_{max}=\qty{21}{\micro m}\simeq 6.6 \sigma$. 
Full video analyzed ($3.1\times10^3$ frames). Inferred active velocity $U\simeq\qty{9.7}{\micro m\per\second}$ and rotational diffusion $D_r \simeq \qty{0.092}{\second^{-1}}$. 

\textit{Set $B$, Fig~\ref{fig:inferred_alexis}}. Parameters for the basis: $N_o=2$, $N_r=16$ and $N_b=285$. The index $l$ for the radial functions $g_l(r)$ spans $l \in \{ 0,1,2,\dots 15 \}$ with $r_0=\qty{1}{\micro m}$. 
Cell dimension for the SFI algorithm: $r_\mathrm{max}=\qty{24}{\micro m} \simeq 7.5 \sigma$. $4\times10^3$ frames analyzed. Inferred active velocity $U\simeq\qty{7.0}{\micro m\per\second}$ and rotational diffusion $D_r \simeq \qty{0.085}{\second^{-1}}$. 

\subsection*{Simulation parameters.}

To simulate the dynamics of the particles, we use the cell-list algorithm with periodic boundary condition.
The dimension of the cell is equal to $r_\mathrm{max}$ given by the SFI analysis.
The inferred forces are activated only for a specific range $r_\mathrm{min}<r<r_\mathrm{max}$ to circumvent the bad estimation at lower distance.
In addition to them, to model the steric repulsion, we systematically add a repulsive interaction deriving from the potential: $V(r)=(\epsilon/2)(1-r/\sigma)^2$ with strength $\epsilon = 800$ and diameter $\sigma = \qty{3.17}{\micro m}$. The number of particles is given by $\phi/(S_\mathrm{box}\pi \sigma^2/4)$ with $\phi$ the area fraction and $S_\mathrm{box}$ the area of the box. 

\textit{Comparison of the observables, Fig~\ref{fig:sim_inferred_daiki}}. For all simulations, we use the same particle number and box size as for the experiment set A. 
Parameters : $r_\mathrm{max}= \qty{21}{\micro m}$, $r_\mathrm{min}= \qty{4}{\micro m}$, time interval $\delta t= \qty{0.1}{s}$ and frame number $\num{4e3}$. The initial distribution of the particles is the same as for the first experimental frame.  

\textit{Inferred simulations with varying area fraction $\phi$, Fig~\ref{fig:flocking}}. Parameters for all simulations : $r_\mathrm{max}=\qty{20}{\micro m}$, $r_\mathrm{min}=\qty{4}{\micro m}$, time interval $\delta t = \qty{0.1}{s}$, frame number $\num{5e3}$ and periodic box of size $420\times\qty{420}{\micro m^2}$. The number of particles depends on the area fraction $\phi$. The initial positions of the the particle are uniformly distributed in the box.

\textit{Evaluation of the pair correlation function $B(\rr)$, Fig~\ref{fig:pcorel_alexis}}. Spatial bin resolution $\qty{0.3}{\micro m}$. For the experiment Fig~\ref{fig:pcorel_alexis}(a), $\num{8e3}$ frames processed with time interval $\delta t = \qty{0.1}{s}$ and average number of particles $\sim 500$. 
For all simulations, $500$ particles constrained within a periodic box of size $360\times \qty{288}{\micro m^2}$, $r_\mathrm{max}=\qty{24}{\micro m}$, $r_\mathrm{min}=\qty{3}{\micro m}$, $\num{8e3}$ frames processed with time interval $\delta t=\qty{0.1}{s}$. 
For the ABP simulation Fig~\ref{fig:pcorel_alexis}(c), the radial interaction is given only by the soft sphere interaction. 

\section*{Conflicts of interest}

There are no conflicts to declare.

\section*{Data availability}

The data that support the findings of this study are available upon reasonable request.

\section*{Acknowledgements}

J. H. and V. D. acknowledge financial support from CNRS through the Emergence@INC program for the project CORRELACT.
P. R. acknowledges funding from the French National Research Agency (ANR-16-CONV-0001), from Excellence Initiative of Aix-Marseille University A*MIDEX and from the European Union (ERC-SuperStoc-101117322).



\balance


\bibliography{biblio_sfi_janus} 

\providecommand*{\mcitethebibliography}{\thebibliography}
\csname @ifundefined\endcsname{endmcitethebibliography}
{\let\endmcitethebibliography\endthebibliography}{}
\begin{mcitethebibliography}{40}
\providecommand*{\natexlab}[1]{#1}
\providecommand*{\mciteSetBstSublistMode}[1]{}
\providecommand*{\mciteSetBstMaxWidthForm}[2]{}
\providecommand*{\mciteBstWouldAddEndPuncttrue}
  {\def\EndOfBibitem{\unskip.}}
\providecommand*{\mciteBstWouldAddEndPunctfalse}
  {\let\EndOfBibitem\relax}
\providecommand*{\mciteSetBstMidEndSepPunct}[3]{}
\providecommand*{\mciteSetBstSublistLabelBeginEnd}[3]{}
\providecommand*{\EndOfBibitem}{}
\mciteSetBstSublistMode{f}
\mciteSetBstMaxWidthForm{subitem}
{(\emph{\alph{mcitesubitemcount}})}
\mciteSetBstSublistLabelBeginEnd{\mcitemaxwidthsubitemform\space}
{\relax}{\relax}

\bibitem[Aranson(2022)]{Aranson2022}
I.~S. Aranson, \emph{Reports on Progress in Physics}, 2022, \textbf{85},
  076601\relax
\mciteBstWouldAddEndPuncttrue
\mciteSetBstMidEndSepPunct{\mcitedefaultmidpunct}
{\mcitedefaultendpunct}{\mcitedefaultseppunct}\relax
\EndOfBibitem
\bibitem[Xu \emph{et~al.}(2023)Xu, Huang, Zhang, and Wu]{Xu2023Autonomous}
H.~Xu, Y.~Huang, R.~Zhang and Y.~Wu, \emph{Nature Physics}, 2023, \textbf{19},
  46--51\relax
\mciteBstWouldAddEndPuncttrue
\mciteSetBstMidEndSepPunct{\mcitedefaultmidpunct}
{\mcitedefaultendpunct}{\mcitedefaultseppunct}\relax
\EndOfBibitem
\bibitem[Vicsek \emph{et~al.}(1995)Vicsek, Czirók, Ben-Jacob, Cohen, and
  Shochet]{Vicsek1995}
T.~Vicsek, A.~Czirók, E.~Ben-Jacob, I.~Cohen and O.~Shochet, \emph{{Phys. Rev.
  Lett.}}, 1995, \textbf{75}, 1226--1229\relax
\mciteBstWouldAddEndPuncttrue
\mciteSetBstMidEndSepPunct{\mcitedefaultmidpunct}
{\mcitedefaultendpunct}{\mcitedefaultseppunct}\relax
\EndOfBibitem
\bibitem[Gautrais \emph{et~al.}(2012)Gautrais, Ginelli, Fournier, Blanco,
  Soria, Chat{\'e}, and Theraulaz]{Gautrais2012}
J.~Gautrais, F.~Ginelli, R.~Fournier, S.~Blanco, M.~Soria, H.~Chat{\'e} and
  G.~Theraulaz, \emph{PLoS Computational Biology}, 2012, \textbf{8},
  e1002678\relax
\mciteBstWouldAddEndPuncttrue
\mciteSetBstMidEndSepPunct{\mcitedefaultmidpunct}
{\mcitedefaultendpunct}{\mcitedefaultseppunct}\relax
\EndOfBibitem
\bibitem[Chen \emph{et~al.}(2023)Chen, Winiarski, Pu{\'s}cian, Knapska,
  Walczak, and Mora]{Chen2023}
X.~Chen, M.~Winiarski, A.~Pu{\'s}cian, E.~Knapska, A.~M. Walczak and T.~Mora,
  \emph{Physical Review X}, 2023, \textbf{13}, 041053\relax
\mciteBstWouldAddEndPuncttrue
\mciteSetBstMidEndSepPunct{\mcitedefaultmidpunct}
{\mcitedefaultendpunct}{\mcitedefaultseppunct}\relax
\EndOfBibitem
\bibitem[Bain and Bartolo(2019)]{Bain2019}
N.~Bain and D.~Bartolo, \emph{{Science}}, 2019, \textbf{363}, 46--49\relax
\mciteBstWouldAddEndPuncttrue
\mciteSetBstMidEndSepPunct{\mcitedefaultmidpunct}
{\mcitedefaultendpunct}{\mcitedefaultseppunct}\relax
\EndOfBibitem
\bibitem[Cordes \emph{et~al.}(2024)Cordes, Schadschneider, and
  Nicolas]{Cordes2024Dimensionless}
J.~Cordes, A.~Schadschneider and A.~Nicolas, \emph{PNAS Nexus}, 2024,
  \textbf{3}, pgae120\relax
\mciteBstWouldAddEndPuncttrue
\mciteSetBstMidEndSepPunct{\mcitedefaultmidpunct}
{\mcitedefaultendpunct}{\mcitedefaultseppunct}\relax
\EndOfBibitem
\bibitem[Gu \emph{et~al.}(2025)Gu, Guiselin, Bain, Zuriguel, and
  Bartolo]{Gu2025}
F.~Gu, B.~Guiselin, N.~Bain, I.~Zuriguel and D.~Bartolo, \emph{Nature}, 2025,
  \textbf{638}, 112--119\relax
\mciteBstWouldAddEndPuncttrue
\mciteSetBstMidEndSepPunct{\mcitedefaultmidpunct}
{\mcitedefaultendpunct}{\mcitedefaultseppunct}\relax
\EndOfBibitem
\bibitem[Bechinger \emph{et~al.}(2016)Bechinger, Di~Leonardo, L{\"o}wen,
  Reichhardt, Volpe, and Volpe]{Bechinger2016}
C.~Bechinger, R.~Di~Leonardo, H.~L{\"o}wen, C.~Reichhardt, G.~Volpe and
  G.~Volpe, \emph{Reviews of Modern Physics}, 2016, \textbf{88}, 045006\relax
\mciteBstWouldAddEndPuncttrue
\mciteSetBstMidEndSepPunct{\mcitedefaultmidpunct}
{\mcitedefaultendpunct}{\mcitedefaultseppunct}\relax
\EndOfBibitem
\bibitem[Deseigne \emph{et~al.}(2010)Deseigne, Dauchot, and
  Chaté]{Deseigne2010}
J.~Deseigne, O.~Dauchot and H.~Chaté, \emph{{Phys. Rev. Lett.}}, 2010,
  \textbf{105}, 098001\relax
\mciteBstWouldAddEndPuncttrue
\mciteSetBstMidEndSepPunct{\mcitedefaultmidpunct}
{\mcitedefaultendpunct}{\mcitedefaultseppunct}\relax
\EndOfBibitem
\bibitem[Weber \emph{et~al.}(2013)Weber, Hanke, Deseigne, Léonard, Dauchot,
  Frey, and Chaté]{Weber2013}
C.~A. Weber, T.~Hanke, J.~Deseigne, S.~Léonard, O.~Dauchot, E.~Frey and
  H.~Chaté, \emph{{Phys. Rev. Lett.}}, 2013, \textbf{110}, 208001\relax
\mciteBstWouldAddEndPuncttrue
\mciteSetBstMidEndSepPunct{\mcitedefaultmidpunct}
{\mcitedefaultendpunct}{\mcitedefaultseppunct}\relax
\EndOfBibitem
\bibitem[Izri \emph{et~al.}(2014)Izri, van~der Linden, Michelin, and
  Dauchot]{Izri2014}
Z.~Izri, M.~N. van~der Linden, S.~Michelin and O.~Dauchot, \emph{{Phys. Rev.
  Lett.}}, 2014, \textbf{113}, 248302\relax
\mciteBstWouldAddEndPuncttrue
\mciteSetBstMidEndSepPunct{\mcitedefaultmidpunct}
{\mcitedefaultendpunct}{\mcitedefaultseppunct}\relax
\EndOfBibitem
\bibitem[Hokmabad \emph{et~al.}(2022)Hokmabad, Agudo-Canalejo, Saha,
  Golestanian, and Maass]{Hokmabad2022Chemotactic}
B.~V. Hokmabad, J.~Agudo-Canalejo, S.~Saha, R.~Golestanian and C.~C. Maass,
  \emph{Proceedings of the National Academy of Sciences}, 2022, \textbf{119},
  e2122269119\relax
\mciteBstWouldAddEndPuncttrue
\mciteSetBstMidEndSepPunct{\mcitedefaultmidpunct}
{\mcitedefaultendpunct}{\mcitedefaultseppunct}\relax
\EndOfBibitem
\bibitem[Lin \emph{et~al.}(2024)Lin, Kim, Arunachalam, Hardian, Adera,
  Aizenberg, Yao, and Daniel]{Lin2024Emergent}
M.~Lin, P.~Kim, S.~Arunachalam, R.~Hardian, S.~Adera, J.~Aizenberg, X.~Yao and
  D.~Daniel, \emph{Phys. Rev. Lett.}, 2024, \textbf{132}, 058203\relax
\mciteBstWouldAddEndPuncttrue
\mciteSetBstMidEndSepPunct{\mcitedefaultmidpunct}
{\mcitedefaultendpunct}{\mcitedefaultseppunct}\relax
\EndOfBibitem
\bibitem[Bricard \emph{et~al.}(2013)Bricard, Caussin, Desreumaux, Dauchot, and
  Bartolo]{Bricard2013}
A.~Bricard, J.-B. Caussin, N.~Desreumaux, O.~Dauchot and D.~Bartolo,
  \emph{{Nature}}, 2013, \textbf{503}, 95--98\relax
\mciteBstWouldAddEndPuncttrue
\mciteSetBstMidEndSepPunct{\mcitedefaultmidpunct}
{\mcitedefaultendpunct}{\mcitedefaultseppunct}\relax
\EndOfBibitem
\bibitem[Palacci \emph{et~al.}(2010)Palacci, Cottin-Bizonne, Ybert, and
  Bocquet]{Palacci2010}
J.~Palacci, C.~Cottin-Bizonne, C.~Ybert and L.~Bocquet, \emph{{Phys. Rev.
  Lett.}}, 2010, \textbf{105}, 088304\relax
\mciteBstWouldAddEndPuncttrue
\mciteSetBstMidEndSepPunct{\mcitedefaultmidpunct}
{\mcitedefaultendpunct}{\mcitedefaultseppunct}\relax
\EndOfBibitem
\bibitem[Bazant and Squires(2010)]{Bazant2010}
M.~Z. Bazant and T.~M. Squires, \emph{Current Opinion in Colloid \& Interface
  Science}, 2010, \textbf{15}, 203--213\relax
\mciteBstWouldAddEndPuncttrue
\mciteSetBstMidEndSepPunct{\mcitedefaultmidpunct}
{\mcitedefaultendpunct}{\mcitedefaultseppunct}\relax
\EndOfBibitem
\bibitem[Boymelgreen \emph{et~al.}(2016)Boymelgreen, Yossifon, and
  Miloh]{Boymelgreen2016}
A.~Boymelgreen, G.~Yossifon and T.~Miloh, \emph{Langmuir : the ACS journal of
  surfaces and colloids}, 2016, \textbf{32}, 9540--9547\relax
\mciteBstWouldAddEndPuncttrue
\mciteSetBstMidEndSepPunct{\mcitedefaultmidpunct}
{\mcitedefaultendpunct}{\mcitedefaultseppunct}\relax
\EndOfBibitem
\bibitem[Nishiguchi and Sano(2015)]{Nishiguchi2015}
D.~Nishiguchi and M.~Sano, \emph{{Phys. Rev. E}}, 2015, \textbf{92},
  052309\relax
\mciteBstWouldAddEndPuncttrue
\mciteSetBstMidEndSepPunct{\mcitedefaultmidpunct}
{\mcitedefaultendpunct}{\mcitedefaultseppunct}\relax
\EndOfBibitem
\bibitem[Yan \emph{et~al.}(2016)Yan, Han, Zhang, Xu, Luijten, and
  Granick]{Yan2016Reconfiguring}
J.~Yan, M.~Han, J.~Zhang, C.~Xu, E.~Luijten and S.~Granick, \emph{{Nature
  Materials}}, 2016, \textbf{15}, 1095 EP --\relax
\mciteBstWouldAddEndPuncttrue
\mciteSetBstMidEndSepPunct{\mcitedefaultmidpunct}
{\mcitedefaultendpunct}{\mcitedefaultseppunct}\relax
\EndOfBibitem
\bibitem[Nishiguchi \emph{et~al.}(2018)Nishiguchi, Iwasawa, Jiang, and
  Sano]{Nishiguchi2018}
D.~Nishiguchi, J.~Iwasawa, H.-R. Jiang and M.~Sano, \emph{{New Journal of
  Physics}}, 2018, \textbf{20}, 015002\relax
\mciteBstWouldAddEndPuncttrue
\mciteSetBstMidEndSepPunct{\mcitedefaultmidpunct}
{\mcitedefaultendpunct}{\mcitedefaultseppunct}\relax
\EndOfBibitem
\bibitem[Zhang \emph{et~al.}(2021)Zhang, Alert, Yan, Wingreen, and
  Granick]{Zhang2021}
J.~Zhang, R.~Alert, J.~Yan, N.~S. Wingreen and S.~Granick, \emph{{Nature
  Physics}}, 2021\relax
\mciteBstWouldAddEndPuncttrue
\mciteSetBstMidEndSepPunct{\mcitedefaultmidpunct}
{\mcitedefaultendpunct}{\mcitedefaultseppunct}\relax
\EndOfBibitem
\bibitem[Poncet \emph{et~al.}(2021)Poncet, Bénichou, Démery, and
  Nishiguchi]{Poncet2021}
A.~Poncet, O.~Bénichou, V.~Démery and D.~Nishiguchi, \emph{{Phys. Rev. E}},
  2021, \textbf{103}, 012605\relax
\mciteBstWouldAddEndPuncttrue
\mciteSetBstMidEndSepPunct{\mcitedefaultmidpunct}
{\mcitedefaultendpunct}{\mcitedefaultseppunct}\relax
\EndOfBibitem
\bibitem[Iwasawa \emph{et~al.}(2021)Iwasawa, Nishiguchi, and Sano]{Iwasawa2021}
J.~Iwasawa, D.~Nishiguchi and M.~Sano, \emph{Phys. Rev. Research}, 2021,
  \textbf{3}, 043104\relax
\mciteBstWouldAddEndPuncttrue
\mciteSetBstMidEndSepPunct{\mcitedefaultmidpunct}
{\mcitedefaultendpunct}{\mcitedefaultseppunct}\relax
\EndOfBibitem
\bibitem[Nishiguchi(2023)]{Nishiguchi2023Deciphering}
D.~Nishiguchi, \emph{Journal of the Physical Society of Japan}, 2023,
  \textbf{92}, 121007\relax
\mciteBstWouldAddEndPuncttrue
\mciteSetBstMidEndSepPunct{\mcitedefaultmidpunct}
{\mcitedefaultendpunct}{\mcitedefaultseppunct}\relax
\EndOfBibitem
\bibitem[Das \emph{et~al.}(2024)Das, Ciarchi, Zhou, Yan, Zhang, and
  Alert]{Das2024Flocking}
S.~Das, M.~Ciarchi, Z.~Zhou, J.~Yan, J.~Zhang and R.~Alert, \emph{Phys. Rev.
  X}, 2024, \textbf{14}, 031008\relax
\mciteBstWouldAddEndPuncttrue
\mciteSetBstMidEndSepPunct{\mcitedefaultmidpunct}
{\mcitedefaultendpunct}{\mcitedefaultseppunct}\relax
\EndOfBibitem
\bibitem[Cichos \emph{et~al.}(2020)Cichos, Gustavsson, Mehlig, and
  Volpe]{Cichos2020}
F.~Cichos, K.~Gustavsson, B.~Mehlig and G.~Volpe, \emph{Nature Machine
  Intelligence}, 2020, \textbf{2}, 94--103\relax
\mciteBstWouldAddEndPuncttrue
\mciteSetBstMidEndSepPunct{\mcitedefaultmidpunct}
{\mcitedefaultendpunct}{\mcitedefaultseppunct}\relax
\EndOfBibitem
\bibitem[Ruiz-Garcia \emph{et~al.}(2024)Ruiz-Garcia, Barriuso~G., Alexander,
  Aarts, Ghiringhelli, and Valeriani]{RuizGarcia2024Discovering}
M.~Ruiz-Garcia, C.~M. Barriuso~G., L.~C. Alexander, D.~G. A.~L. Aarts, L.~M.
  Ghiringhelli and C.~Valeriani, \emph{Phys. Rev. E}, 2024, \textbf{109},
  064611\relax
\mciteBstWouldAddEndPuncttrue
\mciteSetBstMidEndSepPunct{\mcitedefaultmidpunct}
{\mcitedefaultendpunct}{\mcitedefaultseppunct}\relax
\EndOfBibitem
\bibitem[Joshi \emph{et~al.}(2022)Joshi, Ray, Lemma, Varghese, Sharp, Dogic,
  Baskaran, and Hagan]{Joshi2022}
C.~Joshi, S.~Ray, L.~M. Lemma, M.~Varghese, G.~Sharp, Z.~Dogic, A.~Baskaran and
  M.~F. Hagan, \emph{Physical Review Letters}, 2022, \textbf{129}, 258001\relax
\mciteBstWouldAddEndPuncttrue
\mciteSetBstMidEndSepPunct{\mcitedefaultmidpunct}
{\mcitedefaultendpunct}{\mcitedefaultseppunct}\relax
\EndOfBibitem
\bibitem[Supekar \emph{et~al.}(2023)Supekar, Song, Hastewell, Choi, Mietke, and
  Dunkel]{Supekar2023Learning}
R.~Supekar, B.~Song, A.~Hastewell, G.~P.~T. Choi, A.~Mietke and J.~Dunkel,
  \emph{Proceedings of the National Academy of Sciences}, 2023, \textbf{120},
  e2206994120\relax
\mciteBstWouldAddEndPuncttrue
\mciteSetBstMidEndSepPunct{\mcitedefaultmidpunct}
{\mcitedefaultendpunct}{\mcitedefaultseppunct}\relax
\EndOfBibitem
\bibitem[Ha and Jeong(2021)]{Ha2021Unraveling}
S.~Ha and H.~Jeong, \emph{Scientific Reports}, 2021, \textbf{11}, 12804\relax
\mciteBstWouldAddEndPuncttrue
\mciteSetBstMidEndSepPunct{\mcitedefaultmidpunct}
{\mcitedefaultendpunct}{\mcitedefaultseppunct}\relax
\EndOfBibitem
\bibitem[Jung \emph{et~al.}(2017)Jung, Hanke, and Schmid]{Jung2017}
G.~Jung, M.~Hanke and F.~Schmid, \emph{Journal of Chemical Theory and
  Computation}, 2017, \textbf{13}, 2481--2488\relax
\mciteBstWouldAddEndPuncttrue
\mciteSetBstMidEndSepPunct{\mcitedefaultmidpunct}
{\mcitedefaultendpunct}{\mcitedefaultseppunct}\relax
\EndOfBibitem
\bibitem[{Malpica-Morales} \emph{et~al.}(2023){Malpica-Morales}, Yatsyshin,
  {Duran-Olivencia}, and Kalliadasis]{Malpica2023}
A.~{Malpica-Morales}, P.~Yatsyshin, M.~A. {Duran-Olivencia} and S.~Kalliadasis,
  \emph{The Journal of Chemical Physics}, 2023, \textbf{159}, 104109\relax
\mciteBstWouldAddEndPuncttrue
\mciteSetBstMidEndSepPunct{\mcitedefaultmidpunct}
{\mcitedefaultendpunct}{\mcitedefaultseppunct}\relax
\EndOfBibitem
\bibitem[Torquato and Wang(2022)]{Torquato2022}
S.~Torquato and H.~Wang, \emph{Phys. Rev. E}, 2022, \textbf{106}, 044122\relax
\mciteBstWouldAddEndPuncttrue
\mciteSetBstMidEndSepPunct{\mcitedefaultmidpunct}
{\mcitedefaultendpunct}{\mcitedefaultseppunct}\relax
\EndOfBibitem
\bibitem[Frishman and Ronceray(2020)]{Frishman2020}
A.~Frishman and P.~Ronceray, \emph{{Phys. Rev. X}}, 2020, \textbf{10},
  021009\relax
\mciteBstWouldAddEndPuncttrue
\mciteSetBstMidEndSepPunct{\mcitedefaultmidpunct}
{\mcitedefaultendpunct}{\mcitedefaultseppunct}\relax
\EndOfBibitem
\bibitem[Br{\"u}ckner \emph{et~al.}(2020)Br{\"u}ckner, Ronceray, and
  Broedersz]{Bruckner2020}
D.~B. Br{\"u}ckner, P.~Ronceray and C.~P. Broedersz, \emph{Physical Review
  Letters}, 2020, \textbf{125}, 058103\relax
\mciteBstWouldAddEndPuncttrue
\mciteSetBstMidEndSepPunct{\mcitedefaultmidpunct}
{\mcitedefaultendpunct}{\mcitedefaultseppunct}\relax
\EndOfBibitem
\bibitem[Marchetti \emph{et~al.}(2013)Marchetti, Joanny, Ramaswamy, Liverpool,
  Prost, Rao, and Simha]{Marchetti2013}
M.~C. Marchetti, J.~F. Joanny, S.~Ramaswamy, T.~B. Liverpool, J.~Prost, M.~Rao
  and R.~A. Simha, \emph{{Rev. Mod. Phys.}}, 2013, \textbf{85},
  1143--1189\relax
\mciteBstWouldAddEndPuncttrue
\mciteSetBstMidEndSepPunct{\mcitedefaultmidpunct}
{\mcitedefaultendpunct}{\mcitedefaultseppunct}\relax
\EndOfBibitem
\bibitem[Zhang \emph{et~al.}(2017)Zhang, Grzybowski, and Granick]{Zhang2017}
J.~Zhang, B.~A. Grzybowski and S.~Granick, \emph{Langmuir}, 2017, \textbf{33},
  6964--6977\relax
\mciteBstWouldAddEndPuncttrue
\mciteSetBstMidEndSepPunct{\mcitedefaultmidpunct}
{\mcitedefaultendpunct}{\mcitedefaultseppunct}\relax
\EndOfBibitem
\bibitem[Ginot \emph{et~al.}(2015)Ginot, Theurkauff, Levis, Ybert, Bocquet,
  Berthier, and Cottin-Bizonne]{Ginot2015}
F.~Ginot, I.~Theurkauff, D.~Levis, C.~Ybert, L.~Bocquet, L.~Berthier and
  C.~Cottin-Bizonne, \emph{{Phys. Rev. X}}, 2015, \textbf{5}, 011004\relax
\mciteBstWouldAddEndPuncttrue
\mciteSetBstMidEndSepPunct{\mcitedefaultmidpunct}
{\mcitedefaultendpunct}{\mcitedefaultseppunct}\relax
\EndOfBibitem
\bibitem[Narinder \emph{et~al.}(2018)Narinder, Bechinger, and
  Gomez-Solano]{Narinder2018}
N.~Narinder, C.~Bechinger and J.~R. Gomez-Solano, \emph{{Phys. Rev. Lett.}},
  2018, \textbf{121}, 078003\relax
\mciteBstWouldAddEndPuncttrue
\mciteSetBstMidEndSepPunct{\mcitedefaultmidpunct}
{\mcitedefaultendpunct}{\mcitedefaultseppunct}\relax
\EndOfBibitem
\end{mcitethebibliography}
\bibliographystyle{rsc} 

\end{document}